%% file: main.tex
\documentclass[conference]{IEEEtran}
\IEEEoverridecommandlockouts

\usepackage[]{hyperref}

\usepackage{amsmath,amssymb,amsfonts}
\usepackage{amssymb}
\usepackage{textcomp}
\usepackage{multirow}
\usepackage{enumitem}
\usepackage{textcomp}
\usepackage{makecell}
\usepackage{multirow}
\usepackage{tabularray}
\usepackage{array, makecell}

\usepackage{float}
\usepackage{fancyhdr}
\usepackage{colortbl}
\usepackage{arydshln}
\usepackage{graphicx}
\usepackage{textcomp}
\usepackage{soul}
\usepackage{hyperref}
\usepackage{tcolorbox}
\usepackage{wrapfig}
\usepackage{algorithm2e}
\RestyleAlgo{ruled} 
\usepackage{amssymb}
\usepackage{pifont}
\usepackage{amsmath}
\usepackage{graphicx}
\usepackage{textcomp}
\usepackage{makecell}
\usepackage{multirow}
\usepackage{tikz}
\usepackage{float}
\usepackage{graphicx}
\usepackage{booktabs}
\usepackage{cleveref}
\usepackage[margin=0.8in, top=0.8in]{geometry}
\usepackage{tabularx}
\usepackage{xcolor}

\usepackage{caption}

\crefformat{section}{\S#2#1#3} 
\crefformat{subsection}{\S#2#1#3}
\crefformat{subsubsection}{\S#2#1#3}
\usepackage{microtype}
\usepackage{booktabs} 
\usetikzlibrary{shapes,arrows}

\usepackage{pifont}
\usepackage{cite}
\usepackage{amsmath,amsfonts}
\usepackage{newtxtext,newtxmath}
\usepackage{algorithmic}
\usepackage{textcomp}
\usepackage{comment}
\usepackage{fancyhdr}
\usepackage{lipsum}
\usepackage{comment}
\usepackage{outlines}
\usepackage{graphicx}
\usepackage{float}
\usepackage{tablefootnote}
\usepackage[export]{adjustbox}

\usepackage{multirow}
\usepackage{pifont}
\usetikzlibrary{shapes, arrows, positioning}
\usepackage{mdframed}

\usepackage{subcaption}

\usepackage{fontawesome5}
\usepackage{svg}

\usepackage{forest}

\usepackage{pifont}%
\usepackage{comment}
\usepackage{amsmath}

\usepackage{algorithm2e}

\def\BibTeX{{\rm B\kern-.05em{\sc i\kern-.025em b}\kern-.08em
    T\kern-.1667em\lower.7ex\hbox{E}\kern-.125emX}}
\begin{document}
\input{include/shorthands}
\pdfpagewidth=8.5in
\pdfpageheight=11in

\newcommand{\iscasubmissionnumber}{pap744s1}
\newcommand{\ayazdan}[1]{\textcolor{red}{#1}}

\pagenumbering{arabic}

\title{MIST: A Co-Design Framework for Heterogeneous, Multi-Stage LLM Inference}

\author{Abhimanyu Rajeshkumar Bambhaniya\textsuperscript{1,7},  
Hanjiang Wu\textsuperscript{1},
Suvinay Subramanian\textsuperscript{2},
Sudarshan Srinivasan\textsuperscript{3}, \\ 
Souvik Kundu\textsuperscript{4},
Amir Yazdanbakhsh\textsuperscript{5}, 
Midhilesh Elavazhagan\textsuperscript{3}, Madhu Kumar\textsuperscript{3}, \\
Minlan Yu\textsuperscript{6},
Arijit Raychowdhury\textsuperscript{1,7},
Tushar Krishna\textsuperscript{1,7}
\\
  \textsuperscript{1}Georgia Institute of Technology, \textsuperscript{2}Google, \textsuperscript{3}Intel,
  \textsuperscript{4}Intel Labs, 
  \textsuperscript{5}Google DeepMind, \\
    \textsuperscript{6}Harvard University,
    \textsuperscript{7}Infravana
\\\vspace{2mm}
  Corresponding email: \texttt{abambhaniya3@gatech.edu}
}

\maketitle
\thispagestyle{plain}
\pagestyle{plain}


\input{sections/0_sc_abstract.tex}
\input{sections/1_sc_intro}
\input{sections/2_background_new}
\input{sections/3_0_mist}

\input{sections/4_0_implementation}
\input{sections/5_0_eval_isca2026}

\input{sections/6_related_work}

\input{sections/7_conclusion}


\vspace{-1em}
\section*{Acknowledgment}
We have used LLM tools(Claude, Gemini, Grammarly) for proofreading, grammar correction, and continuity checking across sections. All the AI-generated text has been verified by the authors. 
This work was supported in part by CoCoSys, one of seven centers in JUMP 2.0, a Semiconductor Research Corporation (SRC) program sponsored by DARPA, and also by the Georgia Research Alliance based in Atlanta, Georgia under award GRA.25.071.GT.01.a.



\bibliographystyle{IEEEtranS}
\bibliography{refs}
\end{document}

%% file: include/shorthands.tex
\newcommand{\pmark}{$\triangle$}
\newcommand{\cmark}{ \ding{52}}
\newcommand{\xmark}{ 
{\textbf{--}}}
\newcommand{\upmark}{ {\ding{115}}}
\newcommand{\downmark}{ \textcolor{darkred}{\ding{116}}}

\newcommand{\tick}{\includegraphics[height=1.5ex]{figures/checked.png}}
\newcommand{\cross}{\includegraphics[height=1.5ex]{figures/delete.png}}
\newcommand{\exclamation}{\includegraphics[height=1.5ex]{figures/exclamation-2.png}}
\newcommand*\circled[1]{\tikz[baseline=(char.base)]{
            \node[shape=circle,fill,inner sep=0pt] (char) {\textcolor{white}{#1}};}}

\definecolor{revA}{HTML}{000000}
\definecolor{revB}{HTML}{000000}
\definecolor{revC}{HTML}{000000}
\definecolor{revD}{HTML}{000000}
\definecolor{revE}{HTML}{000000}
\definecolor{revF}{HTML}{000000}
\definecolor{colorRevA}{rgb}{1, 0.90, 0.80}

\definecolor{colorRevB}{rgb}{0.97, 0.80, 0.796}
\definecolor{colorRevC}{rgb}{0.87, 0.83, 0.90}
\definecolor{colorRevD}{rgb}{0.83, 0.91, 0.83}
\definecolor{colorRevE}{rgb}{0.85, 0.90, 0.98}

\definecolor{colorRevF}{rgb}{0.80, 0.93, 0.95}
\definecolor{blond}{rgb}{0.98, 0.94, 0.75}

\newcommand{\colorciteB}[1]{\tcbox[colframe=colorRevB, colback=colorRevB, boxrule=0mm, arc=0mm, left=0mm, right=0mm, top=0mm, bottom=0mm, boxsep=0pt, on line]{\cite{#1}}}

\newcommand{\colorrefA}[1]{\tcbox[colframe=colorRevA, colback=colorRevA, boxrule=0mm, arc=0mm, left=0mm, right=0mm, top=0mm, bottom=0mm, boxsep=0pt, on line]{\cref{#1}}}

\newcommand{\colorciteA}[1]{\tcbox[colframe=colorRevA, colback=colorRevA, boxrule=0mm, arc=0mm, left=0mm, right=0mm, top=0mm, bottom=0mm, boxsep=0pt, on line]{\cite{#1}}}

\DeclareRobustCommand{\RevA}[1]{{\sethlcolor{colorRevA}\hl{#1}}}
\DeclareRobustCommand{\RevB}[1]{{\sethlcolor{colorRevB}\hl{#1}}}
\DeclareRobustCommand{\RevC}[1]{{\sethlcolor{colorRevC}\hl{#1}}}
\DeclareRobustCommand{\RevD}[1]{{\sethlcolor{colorRevD}\hl{#1}}}
\DeclareRobustCommand{\RevE}[1]{{\sethlcolor{colorRevE}\hl{#1}}}
\DeclareRobustCommand{\RevF}[1]{{\sethlcolor{colorRevF}\hl{#1}}}
\DeclareRobustCommand{\RevCom}[1]{{\sethlcolor{blond}\hl{#1}}}

\newcommand{\titleRev}[2]{
    \texorpdfstring{{\sethlcolor{#2}\hl{#1}}}{#1}
}

\newcommand{\AB}[1]{{\color{magenta}\bfseries [Abhi:: #1]}}
\newcommand{\TK}[1]{{\color{cyan}\bfseries [Tushar:: #1]}}
\newcommand{\Suv}[1]{{\color{brown}\bfseries [Suvinay:: #1]}}
\newcommand{\SK}[1]{{\color{magenta}\bfseries [Souvik:: #1]}} 
\newcommand{\Sud}[1]{{\color{cyan}\bfseries [Sudarshan:: #1]}} 
\newcommand{\HW}[1]{{\color{violet}\bfseries [Hanjiang:: #1]}}
\newcommand{\todo}[1]{\textcolor{red}{TODO:: #1}}
\newcommand{\rev}[1]{\textcolor{red}{#1}}

\def\Snospace~{\S{}}
\renewcommand*\sectionautorefname{\Snospace}
\def\sectionautorefname{Sec.}
\def\subsectionautorefname{Sec.}
\def\subsubsectionautorefname{Sec.}

\newcommand{\circleNumber}[1]{
{\large \textcircled{\small #1}} }
\newcommand{\SubItem}[1]{
    {\setlength\itemindent{15pt} \item[-] #1}
}
\newcommand{\abc}{{\sf abc}\xspace}
\newcommand{\dse}{{\sf dse}\xspace}
\newcommand{\defns}{{\sf def}}
\newcommand{\tool}{{MIST }}
\newcommand{\toolns}{{MIST}}

\newcommand{\niparagraph}[1]{\vspace{1pt}\noindent\textbf{#1}}
\newcommand{\checkbox}{\textbf{\textcolor{green}{\ding{51}}}}
\newcommand{\xbox}{\textbf{\textcolor{red}{\ding{55}}}}

\newcommand*\hcircled[1]{\tikz[baseline=(char.base)]{\node[shape=circle,draw,inner sep=2pt] (char) {#1};}}

\def\figureautorefname{Fig.}

\definecolor{turqoise}{HTML}{4BCAD2}
\definecolor{orangered}{HTML}{CF1040}
\definecolor{redgreen}{HTML}{66AA00}

\newcommand{\squishlist}{
 \begin{list}{$\bullet$}
  { \setlength{\itemsep}{0pt}
     \setlength{\parsep}{3pt}
     \setlength{\topsep}{3pt}
     \setlength{\partopsep}{0pt}
     \setlength{\leftmargin}{1.5em}
     \setlength{\labelwidth}{1em}
     \setlength{\labelsep}{0.5em} } }

\newcommand{\squishlisttwo}{
 \begin{list}{$\bullet$}
  { \setlength{\itemsep}{0pt}
     \setlength{\parsep}{0pt}
    \setlength{\topsep}{0pt}
    \setlength{\partopsep}{0pt}
    \setlength{\leftmargin}{2em}
    \setlength{\labelwidth}{1.5em}
    \setlength{\labelsep}{0.5em} } }

\newcommand{\squishend}{
  \end{list}  }

\newtcolorbox{boxA}{
    fontupper = \it,
     borderline = {1pt}{0pt}{main, dashed} 
}

%% file: sections/0_sc_abstract.tex
\begin{abstract}
Modern LLM serving now spans multi-stage pipelines including RAG retrieval and KV cache reuse, each with distinct compute, memory, and latency demands.
Inference engines expose a large configuration space with no systematic navigation methodology, and exhaustively benchmarking configurations can exceed \$40K in cloud costs.
Simultaneously, the hardware landscape is rapidly diversifying across AMD GPUs, TPUs, and custom ASICs, while cross-vendor prefill-decode (PD) disaggregated configurations lack unified software stacks for end-to-end evaluation today.

We present \tool\footnote{\textbf{\underline{M}}ulti-stage AI \textbf{\underline{I}}nference \textbf{\underline{S}}imulation \textbf{\underline{T}}oolkit}, an event-driven, end-to-end simulation framework addressing all three challenges.
\tool models the full serving stack across an \textbf{AI Workload Layer}, a \textbf{System \& Software Layer}, and a \textbf{Hardware Layer}, with pluggable accelerator models supporting cross-vendor evaluation without a unified software stack.
\tool achieves end-to-end fidelity within 6\% of real deployed systems.
Across representative workloads, multi-vendor PD-disaggregated deployments optimized with \tool yield up to 49.3\% higher tokens/\$ than single-vendor homogeneous PD disaggregation.
\tool further enables KV storage co-design, revealing how memory hierarchy choices shape tail latency across workloads with varying KV reuse patterns.
\end{abstract}

%% file: sections/1_sc_intro.tex
\section{Introduction}
\label{sec:intro}

Large Language Model (LLM) inference has rapidly become a principal workload in modern computing infrastructure.
Inference must satisfy strict latency, throughput, and cost constraints under highly variable request patterns, spanning trillion-parameter frontier models for complex reasoning~\cite{hao_reasoning_2023, wei2023cot} and agentic workflows~\cite{Lin2025Parrot,yang2024sweagent} to compact models powering chatbots and retrieval-based assistants~\cite{chatgpt, Gemini, copilot}.
As adoption accelerates, the efficiency of LLM \emph{serving} platforms increasingly determines the usable performance of AI systems at scale, and navigating their design space has become a challenge in its own right.

\input{figures/tex/overview_fig}
\input{figures/tex/request_examples}

Today's LLM inference is mediated by \emph{inference serving engines} such as vLLM~\cite{kwon2023efficient}, SGLang~\cite{zheng2024sglang}, llama.cpp~\cite{Ggerganov}, TensorRT-LLM~\cite{TensorRT-LLM}, and NVIDIA Triton~\cite{triton}, which sit between AI applications and the underlying hardware.
As shown in \autoref{fig:overview_figure}(a), an inference server exposes a large number of configurable \emph{knobs}: AI model selection and routing policy; request scheduling (prefix cache placement); token generation batching (continuous, chunked, mixed, disaggregated); inter-node communication (KV cache transfer, context retrieval); kernel optimizations (parallelism, flash attention, KV memory organization); and model-level optimizations (sparsity, quantization, KV compression).
Unfortunately, there lacks a systematic methodology for navigating these knobs today. Practitioners rely on community wisdom and expensive trial-and-error to configure these knobs for deployments.
%
As an example, exhaustively benchmarking Llama3-70B configurations for a single H100 8-GPU deployment would exceed \$40{,}000 in cloud cost.
%
\textit{\textbf{
Challenge 1: The configuration space of LLM inference engines is vast and interdependent, with no systematic methodology to identify optimal deployments - making exhaustive search prohibitively expensive.
}}

Beyond configuration complexity, LLM serving is no longer just prefill and decode: a news search adds a RAG lookup, and code generation adds KV cache retrieval. This is shown in \autoref{fig:req_examples}(a).
These additional stages may not always dominate end-to-end runtime, but they introduce non-trivial secondary effects that cannot be ignored.
RAG retrieval injects a large external context that must be transferred from the retrieval device to the prefill accelerator, increasing effective prompt length and Time-To-First-Token (TTFT).
KV cache retrieval transfers previously stored key-value tensors from tiered storage (DRAM, SSD, or remote memory) to the prefill and decode devices, adding memory bandwidth pressure and transfer latency on top of compute savings.
In disaggregated deployments, these cross-device data movements compound with stage-to-stage handoffs, making the network and memory fabric a first-order performance concern.
\textit{\textbf{Challenge 2: Modern AI requests are multi-stage; even when individual stage runtimes are modest, the associated data transfers and their downstream effects on context size and KV state make end-to-end modeling essential.}}

This multi-stage complexity is further compounded by a rapidly diversifying hardware ecosystem beyond NVIDIA GPUs.
AMD GPUs offer competitive compute density, Google TPUs provide high memory bandwidth at lower cost, and custom ASICs (e.g., Trainium, Inferentia, Etched) target specific patterns such as attention-heavy decode or memory-bound prefill.
Emerging paradigms such as prefill-decode (PD) disaggregation further expand the design space by enabling different hardware to handle different inference phases.
Even for a simple news search pipeline (Pre $\to$ RAG $\to$ Prefill $\to$ Decode $\to$ Post), the same hardware nodes can be mapped in many ways: all stages on a single node, RAG on a CPU while prefill and decode share a GPU, or each stage on a dedicated node (\autoref{fig:req_examples}(b)).
Recent industry developments like NVIDIA$+$Groq~\cite{nvidia_groq_lpx}, AWS$+$Cerebras~\cite{aws_cerebras_2026}, and Gimlet Labs' heterogeneous inference cloud~\cite{gimlet_labs_2026} highlight the growing adoption of multi-vendor hardware for LLM inference.
Most such hardware combinations lack a mature software stack and cannot be benchmarked end-to-end today, forcing architects to make multi-year investment decisions without evaluating cross-vendor pipelines at scale.
\textit{\textbf{Challenge 3: Mapping diverse inference stages to heterogeneous multi-vendor hardware, and reasoning about interactions across stages and chip boundaries, creates a non-trivial challenge for efficient LLM inference deployment.}}

We present \toolns, an event-driven, end-to-end simulation framework for heterogeneous, multi-stage AI inference pipelines that addresses all three challenges.
\tool models the full serving stack across an \textbf{AI Workload Layer} (diverse pipeline stages), a \textbf{System \& Software Layer} (scheduling, batching, and kernel optimizations), and a \textbf{Hardware Layer} (logical abstraction of compute, memory, and network resources), with pluggable accelerator models enabling evaluation of cross-vendor deployments without a unified software stack.
\tool provides:
\begin{itemize} [leftmargin=0.8em]
    \item Diverse inference stages (RAG, KV cache retrieval across DRAM/SSD/remote memory, reasoning, prefill, and decode) with multi-model pipelines and heterogeneous PD-disaggregated configurations.
    \item Global routing and load-balancing; client-level scheduling; batching strategies; and KV cache management (placement, eviction, reuse) across heterogeneous clients and memory tiers.
    \item Modular hardware abstractions supporting real-HW plug-ins, runtime predictors, or cycle-accurate simulators for accelerators not yet in production.
\end{itemize}
Across validated hardware platforms, \tool achieves end-to-end fidelity within 6\% of real-system performance.

Using \tool, we answer two classes of previously inaccessible questions:
\begin{itemize}[leftmargin=0.8em]
    \item \textbf{Heterogeneous deployment optimization:} Across multi-stage workloads, heterogeneous multi-vendor PD-disaggregated configurations outperform homogeneous baselines, and \tool-optimized deployments yield up to 49.3\% tokens/\$ over baselines (\autoref{sec:llm_engine_optimization}).
    \item \textbf{KV storage architecture co-design:} \tool reveals how memory-hierarchy design, and cache granularity interact with end-to-end latency across workloads with varying KV length and reuse patterns (\autoref{subsec:c6_kv_storage}).
\end{itemize}
\tool provides the first end-to-end framework spanning heterogeneous hardware, multi-stage pipelines, and KV storage co-design, offering a principled foundation for optimizing current deployments and architecting next-generation AI infrastructure.

%% file: figures/tex/overview_fig.tex
\begin{figure*}[!bthp]
    \centering
    \includegraphics[width=1\linewidth]{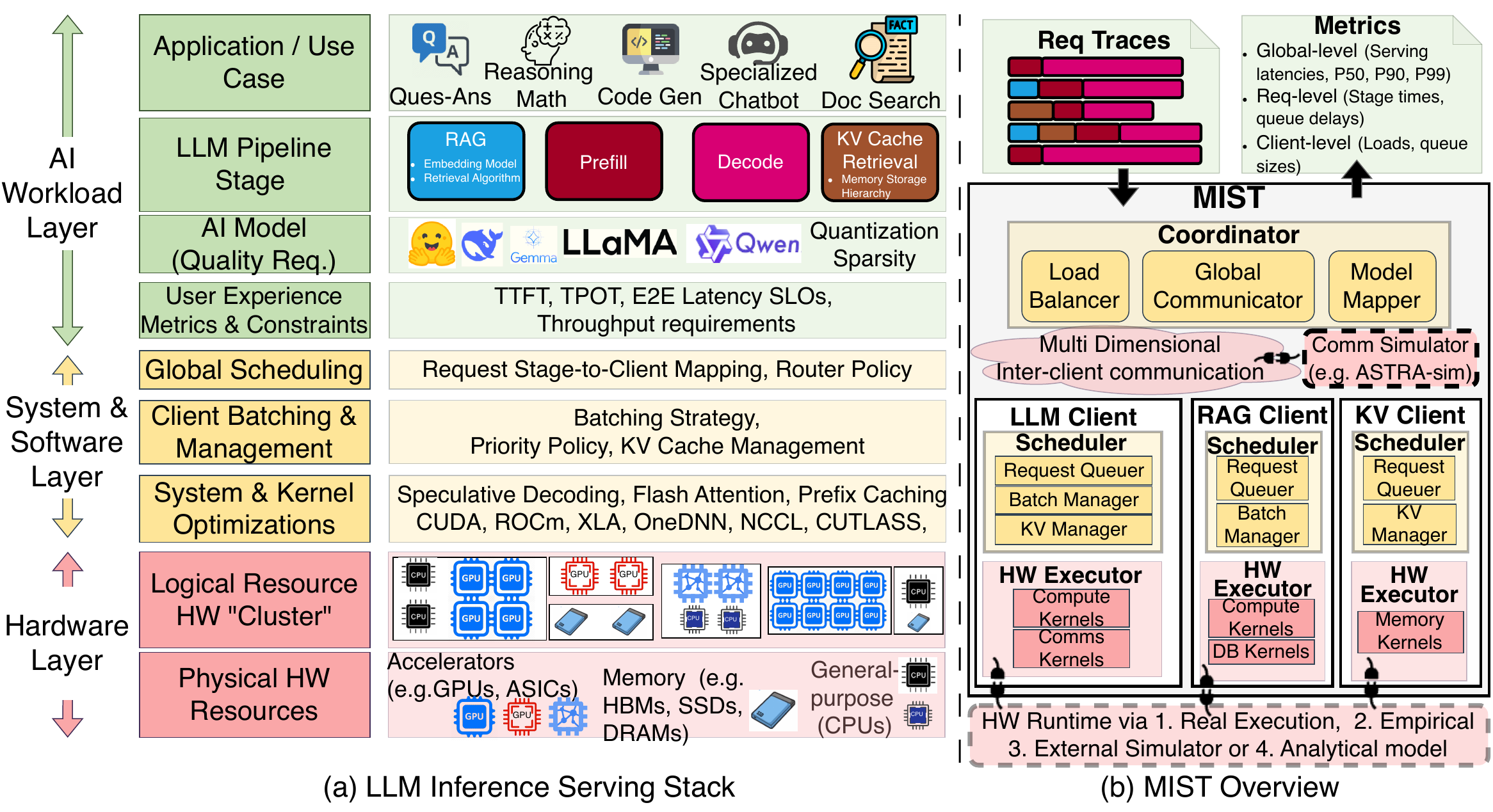}        \vspace{-2em}
    \caption{Understanding the LLM inference serving stack and how \tool models the stack. \tool simulates a collection of clients. Each client consists of a scheduler that issues steps (e.g., tokens for decode, chunks for prefill, Embedding/Retrieval for RAG) to an HW cluster (e.g., Nvidia HGX, AMD MI300X, CPU host with an offloading memory instance, etc.). The HW cluster is a collection of multiple HW Nodes (e.g., GPUs, ASICs, memory, and CPUs).}
    \label{fig:overview_figure}
    \vspace{-2em}
\end{figure*}

%% file: figures/tex/request_examples.tex
\begin{figure}[!t]
    \centering
    \includegraphics[width=\linewidth]{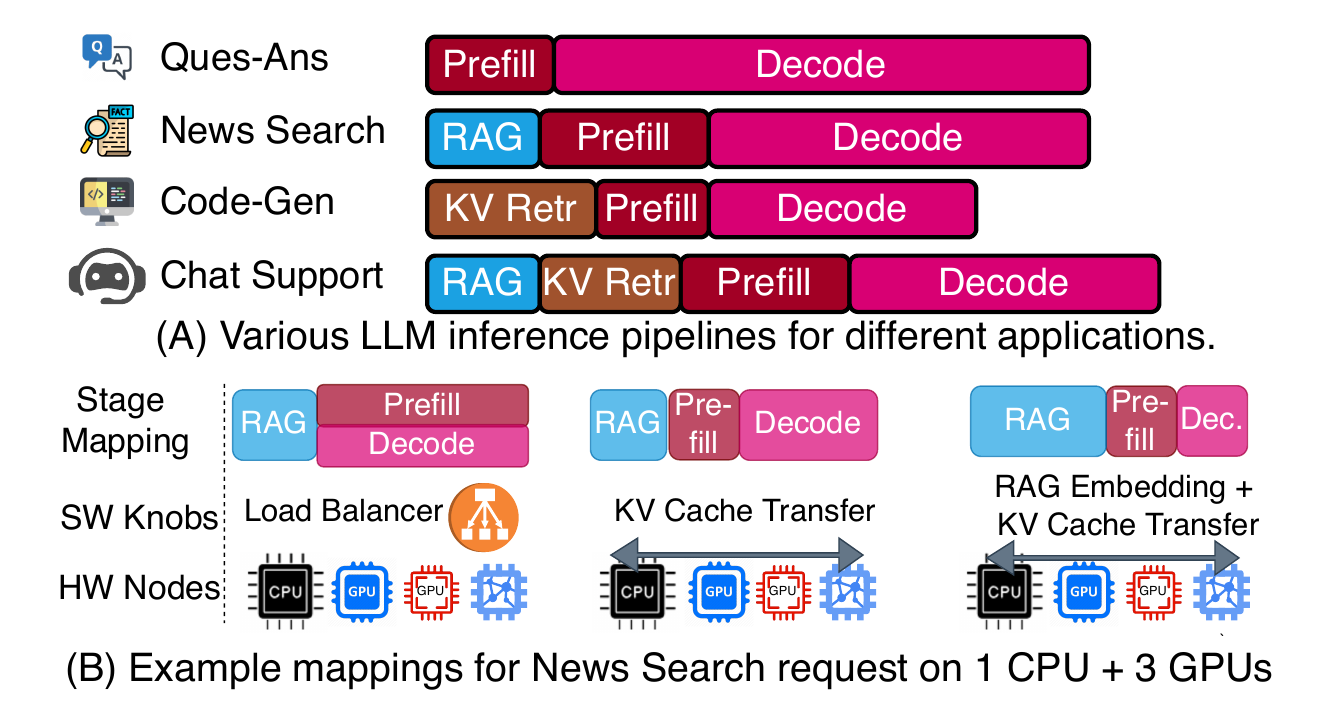}
    \vspace{-1.5 em}
    \caption{(a) LLM inference request types: \emph{Question-answering} (Standard); \emph{News search} (RAG pipeline)~\cite{gao2024ragsurvey}; \emph{Code generation} (KV cache reuse)~\cite{jiang2024codegen}; \emph{Chat support} (RAG + KV cache)~\cite{dam2024chatbotsurvey} (b) Different ways of mapping news search requests across 1 CPU and 3 heterogeneous GPU nodes. For each mapping different sw knobs like load balancer, global communicator need to be activated.
}
    \vspace{-2em}
    \label{fig:req_examples}
\end{figure}

%% file: sections/2_background_new.tex
\section{Background on LLM Inference Pipelines}
\label{sec:background}

\autoref{fig:req_examples} showcases the components within the workflow of modern inference.
Diverse application use cases demand different pipelines. For example, a news search involves a RAG lookup followed by an auto-regressive prefill and decode while accurately solving complex mathematical problems typically interleaves generation with reward-model evaluation~\cite{ahn2024llmmaths,lightman2023letsverify}.
\vspace{-0.75mm}

\subsection{Stages of LLM Inference}
\label{subsec:stages}

A modern LLM inference request passes through several distinct computational stages, each with different resource demands and latency characteristics.

\niparagraph{Prefill and Decode.}
\textbf{Prefill} performs a single forward pass over the input prompts to generate the first token, making it compute-intensive. \textbf{Decode} then proceeds auto-regressively, generating tokens sequentially. This stage is memory-bound and benefits from batching multiple tokens for higher throughput.
Chunked (Aggregated) batching and Disaggregated batching are the most popular ways of batching prefill and decode for concurrent requests to improve the efficiency and throughput of the system.

\textbf{Chunked Batching} improves system throughput by splitting long input sequences into smaller, fixed-size chunks.

\textbf{Disaggregated Batching} decouples prefill and decode stages by assigning them to independently scaled hardware instances, enabling flexible resource allocation for heterogeneous workloads.
Splitwise~\cite{patel2023splitwise} introduced the idea of heterogeneous GPU use, but the software stack is still evolving, and multi-vendor disaggregated LLM inference lacks open-source support. Recent trends in the industry~\cite{nvidia_groq_lpx, aws_cerebras_2026, gimlet_labs_2026} highlight the growing adoption of \textbf{multi-vendor hardware} for disaggregated inference.

\niparagraph{Retrieval Augmented Generation (RAG).}
As shown in \autoref{fig:rag_pipeline}(a), RAG retrieves relevant documents via embedding models and ANN search (e.g., FAISS~\cite{johnson2017billionscalesimilaritysearchgpus}) before augmenting the prompt for LLM generation, improving factuality and reducing hallucination. We adopt DPR(Dense Passage Retrieval)~\cite{karpukhin2020densepassageretrievalopendomain} for embedding and IVF-PQ~\cite{nvidia_ivf_pq} as our ANN method for its balance of memory efficiency and recall at billion-scale~\cite{Chooseth12}.

\input{figures/tex/RAG_pipeline}

\niparagraph{KV Cache Retrieval.}
As shown in \autoref{fig:rag_pipeline}(b), Prefix Caching~\cite{vllm_apc} reduces TTFT by reusing KV cache entries from prior queries that share a prefix, bypassing redundant prefill computation~\cite{yao2024cacheblend, cheng2024large, liu2024cachegen}. KV caches can also be persisted across sessions to support multi-turn conversations~\cite{yao2024cacheblend}.

\subsection{Architecture of Inference Serving Server}
\label{subsec:serving-arch}

Table~\ref{tab:llm-inference} shows the module hierarchy of a modern LLM inference server. The top-level \textbf{Inference Server} owns a \emph{Coordinator} which orchestrates \emph{Load Balancer}, \emph{Global Communicator} (e.g., for inter-client KV cache transfer), and \emph{Model Mapper}. Within each \textbf{Inference Client}, a \textbf{Scheduler} handles request ordering, batching strategy, and KV cache management, while \textbf{HW Executor} has accelerator-specific optimized kernels for compute (e.g., CUDA/ROCm/XLA) and communication (e.g., NCCL/RCCL).

\input{tables/llm_inference_server}

%% file: figures/tex/RAG_pipeline.tex
\begin{figure}[!btp]
    \centering
    \includegraphics[width=1\linewidth]{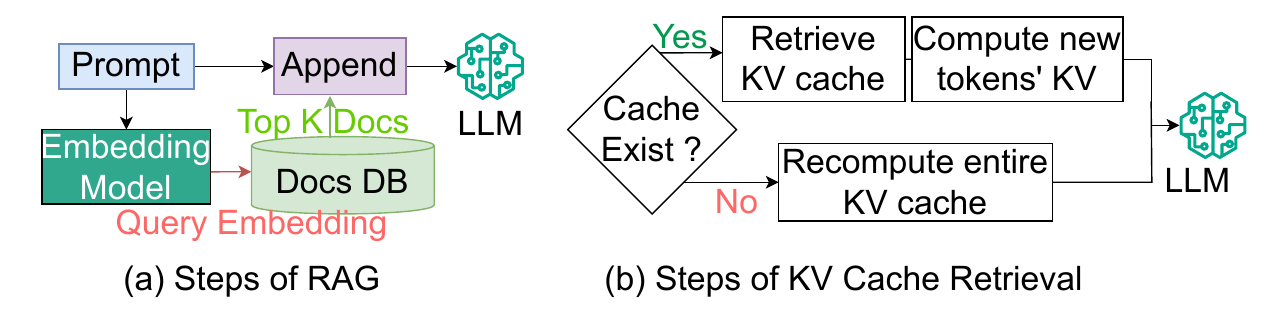}
    \vspace{-1.5em}
    \caption{Steps in (a) RAG \& (b) KV Retrieval.}
    \vspace{-1em}
    \label{fig:rag_pipeline}
\end{figure}

%% file: tables/llm_inference_server.tex

\definecolor{headerblue}{RGB}{30,80,140}
\definecolor{l0bg}{RGB}{213,228,244}
\definecolor{l1bg}{RGB}{232,243,255}
\definecolor{l2bg}{RGB}{225,242,235}
\definecolor{l3bg}{RGB}{240,248,244}
\definecolor{l4bg}{RGB}{250,250,248}
\definecolor{SW_layer_table1}{RGB}{250,230,162}
\definecolor{HW_layer_table1}{RGB}{238,171,170}
\definecolor{SW_layer_table1_lite}{RGB}{253,247,226}
\definecolor{HW_layer_table1_lite}{RGB}{249,229,229}

\definecolor{rulecol}{RGB}{160,185,210}
\definecolor{textdark}{RGB}{25,40,60}

\renewcommand{\arraystretch}{1.4}

\newcommand{\crule}{\arrayrulecolor{rulecol}\midrule\arrayrulecolor{black}}
\newcommand{\li}{\hspace{0.5em}}
\newcommand{\lii}{\hspace{1.6em}}
\newcommand{\liii}{\hspace{2.4em}}


\begin{table}[ht]
\centering
\caption{Modules of \textbf{LLM Inference Server}.}
\label{tab:llm-inference}

\begin{tabularx}{\columnwidth}{
    >{\raggedright\arraybackslash}p{3cm}
    >{\raggedright\arraybackslash}X
}
\rowcolor{headerblue}
\textcolor{white}{\textbf{Module}} &
\textcolor{white}{\textbf{Description}} \\

\rowcolor{l0bg}
\textcolor{textdark}{\textbf{Inference Server}} &
Receives req \& sends response.\\
\rowcolor{SW_layer_table1_lite}
\li Coordinator &
Manages coordination between clients.\\
\rowcolor{SW_layer_table1}
\lii Load Balancer &
Decodes request$\mapsto$client mapping .\\
\rowcolor{SW_layer_table1}
\lii Global Communicator &
Inter client comm(e.g. KV cache $\Leftrightarrow$).\\
\rowcolor{SW_layer_table1}
\lii Model Mapper &
Routes request to appropriate LLM.\\

\specialrule{0.08em}{0pt}{0pt}
\rowcolor{l1bg}
\li\textbf{Inference Client} &
Hosts a single LLM replica \\
\specialrule{0.08em}{0pt}{0pt}
\rowcolor{SW_layer_table1_lite}
\lii\textit{Scheduler} &
Selects requests for each fwd pass.\\ 
\rowcolor{SW_layer_table1}
\liii Request Queuer &
Orders requests by packing policy.\\ 
\rowcolor{SW_layer_table1}
\liii Batch Manager &
Batch requests by batching policy. \\
\rowcolor{SW_layer_table1}
\liii KV Manager &
Manages KV Cache across GPU, CPU, and tiered KV caches. \\
\specialrule{0.08em}{0pt}{0pt}

\rowcolor{HW_layer_table1_lite}
\lii\textit{HW Executor} &
HW specific model executor \\
\rowcolor{HW_layer_table1}
\liii Compute Kernels &
Optimized kernels for the target HW. \\
\rowcolor{HW_layer_table1}
\liii Comms Kernels &
Collective kernels for parallelism. \\

\bottomrule
\end{tabularx}
\vspace{-1em}
\end{table}


%% file: sections/3_0_mist.tex

\section{\textbf{MIST}: \textbf{M}ulti-stage AI \textbf{I}nference \textbf{S}imulation \textbf{T}oolkit}
\label{sec:mist}


We introduce \toolns, a simulation framework designed to capture the complexity of real-world LLM inference pipelines. To resemble real deployments, \tool simulates the end-to-end execution of state-of-the-art LLM inference pipelines across the three layers of the LLM Inference serving stack shown in \autoref{fig:overview_figure}(a). 

%

\input{sections/3_1_overview}
\input{sections/3_2_workload-layer}
\input{sections/3_3_system-layer}
\input{sections/3_4_hw-layer}

\input{sections/3_5_outputs}

%% file: sections/3_1_overview.tex
\subsection{Overview}\label{subsec:hermes_sim_methodology}





\autoref{fig:overview_figure}(b) shows an overview of \tool.
The \textbf{AI Workload Layer} knobs are the input to the \tool framework. Use case specific \emph{requests} are injected into the framework at different arrival times. A \textbf{Request} (\autoref{subsec:request_modelling}) is a pipeline of \emph{stages}, where each stage could have one or more \emph{steps} - for instance, RAG has model embedding and document retrieval (\autoref{fig:rag_pipeline}).

The \textbf{System and Software Layer} forms the core of \toolns. The flow starts with a \textbf{Coordinator} (\autoref{subsec:coordinator}) that performs global scheduling of the inference pipeline. Specifically, it maps and routes different stages of requests to appropriate \emph{clients} based on routing and load-balancing policies. Each \textbf{Client} (\autoref{subsec:clients}) employs a runtime \textbf{Scheduler} (\autoref{subsec:Schedulers}) for executing the different steps of the mapped stage over its underlying HW \emph{cluster} using its \textbf{HW Executor}.



The \textbf{HW Layer} is modeled as \textbf{Cluster} (\autoref{subsec:cluster_modeling}), a logical collection of the hardware nodes within a physical AI platform 
\footnote{For e.g., a physical HGX box with 8×H100 GPUs and 2xCPUs can be mapped as two logical clusters, each with 4xH100+1CPU,  where the first cluster is made a prefill client and the second is a decode client in a disaggregated setup.
}. 
The hardware nodes in the cluster can be any viable combination of GPUs, ASICs, CPUs, and memories, along with their connectivity.
For each client's HW executor runtime simulation, users can choose between (i) real execution runtime, (ii) empirical observed runtime values from real HW, (iii) analytical modeling equation, or (iv) external simulators. 

Thus, each client incorporating a scheduler and an HW executor is similar to real-world LLM frameworks such as SGLang \cite{zheng2024sglang} and vLLM \cite{kwon2023efficient}.

The final outputs (\autoref{subsubsec:output_metrics}) of \tool are various LLM performance metrics at the request level, client level, and coordinator level, like latency, throughput, etc.
 



%% file: sections/3_2_workload-layer.tex
\subsection{Workload Layer: Inputs and Parameters}
\tool accepts request traces as input from AI workload layer. 
In the system and software layer, the router policy, client configuration (Scheduler config + HW clusters), and global client connections are user-defined hyperparameters.

\subsubsection{Request Modeling}
\label{subsec:request_modelling}

\tool currently supports requests that are any combination of four stages of the LLM pipeline
(\textbf{\emph{RAG}}, \textbf{\emph{Prefill}},  \textbf{\emph{Decode}}, and \textbf{\emph{KV Cache Retrieval}}). 
Users can construct an arbitrary combination of stages, as shown in \autoref{fig:req_examples}, to simulate a modern AI use case\footnote {Stages can be extended to model tool calls in agentic workflows.}.

\subsubsection{Input Datasets and Workloads}

Inference begins with feeding a trace of requests into the system. 

\textbf{Request size}: To model diverse prefill and decode token workloads, we use a combination of real and synthetic traces.\textit{Real traces} from production services, such as Azure trace~\cite{azurellmtrace2023} (Conv and Code), that capture realistic input-output token distributions. 
\textit{Synthetic traces} are generated based on observed characteristics in common workloads. They are modeled as a normal distribution with a user-configurable mean and variance for input and output tokens.

\textbf{Request injection} is modeled using a range of models including uniform, normal, poisson, and bursty distributions. This approach better reflects real-world traffic patterns and enables more robust evaluation of system behavior under diverse operational scenarios.

Additionally, each request includes additional parameters depending on the associated stages (e.g., RAG requests require RAG algorithm parameters). 

%% file: sections/3_3_system-layer.tex
\input{algorithms/coordinator_sim}

\subsection{System SW Layer: Coordinator}
\vspace{-0.5em}
\label{subsec:coordinator}
The coordinator manages end-to-end inference execution across clients, ensuring ordered stage scheduling and inter-stage communication. 
\autoref{algo:coordinator_sim} shows the core simulation loop, which integrates event scheduling (\textit{Client} \& \textit{Request} events), routing, and inter-client communication in a unified discrete-event framework.
\textit{Request event:} New requests entering the system or a client returns a request after servicing a stage.
\textit{Client event:} Scheduling + Execution of the steps for requests assigned to a client. E.g., for a prefill/decode client running on an 8xH100 GPU with chunked batching, the client event schedules a single chunk batch with assigned prefill and decode batches, and simulates the HW runtime of the scheduled batch.
\textit{Client Transfer event:} Communication from one client to another based on how the stages are assigned to different clients. E.g., for disaggregated batching, this would be KV cache transfer from prefill to decode client.

%


\subsubsection{Load Balancer}
To determine the next client for a given request stage, the coordinator uses a routing module. 
When multiple clients can execute the same stage, the router uses the user-defined load-balancing policy to distribute work efficiently. 
%
%
We support four routing policies:
Round Robin,
Least work outstanding,
Load-based,
Heavy-Light load split ~\cite{jain2025intelligentrouterllmworkloads}.

Load in the latter two policies can be defined using various request attributes, such as: i) input context length,
ii) current KV cache size (would matter if the HW cluster has different memory capacity),
iii) tokens remaining to be generated\footnote{We do not predict number of output tokens, nor do we use it for any of our experiments. There is existing research~\cite{jain2025intelligentrouterllmworkloads} which can be used to predict number of output tokens.}.
These metrics enable up to nine distinct routing strategies. \tool has a highly modular router API allowing new routing policies to be integrated with minimal effort.   

\subsubsection{Model Mapper} For different AI model instantiations, each request contains metadata specifying the target model.
\footnote{Future extensions may support adaptive model routing based on request characteristics such as complexity, quality, or priority.}
The router can also exploit client placement information to reduce communication costs, especially in disaggregated serving, where large KV caches must be transferred between clients.

\subsubsection{Global Communicator}

Once a routing decision is made, the global communication simulator handles data transfers between clients.
It estimates communication overhead based on data size and transfer granularity (e.g., full KV cache vs. layer-wise transfer~\cite{patel2023splitwise}), accounting for transitions between the multi-dimensional network hierarchy of HW nodes.
%
For simulating multi-dimensional inter-client communication, \tool integrates with ASTRA-sim\footnote{Astra-Sim is a network simulator that can model multi-dimensional networks in a contention-aware fashion~\cite{astrasim_doc_ns3}.}~\cite{astrasim}, enabling accurate modeling of communication latency and bandwidth constraints.
After the data is transferred, the target client can start the stage processing.

%

\vspace{-0.5em}
\subsection{System and Software Layer: Clients}
\label{subsec:clients}
\vspace{-0.4em}

\tool has three different client types: \textit{LLM Client(for Prefill and Decode)}, \textit{RAG Client}, and \textit{KV Retrieval Client} covering most modern LLM use-cases. 
Each client in \tool is composed of a runtime Scheduler 
and a HW Executor. Drawing inspiration from vLLM~\cite{kwon2023efficient}, each client operates at the granularity of a step. In the case of prefill and decode, a step corresponds to a single forward pass. During the execution of a step, new requests may arrive asynchronously but cannot preempt the ongoing computation. 
%
With the scheduled batch at each step, the runtime and cost are modeled separately for each client type. For example, KV cache retrieval depends on both the size of the KV cache and the details of the hierarchical memory architecture to estimate fetch latencies. Further details on how different hardware clusters are simulated according to different client types are provided in \autoref{subsec:cluster_modeling}.

\vspace{-1mm}
\subsection{System SW Layer: Schedulers}
\label{subsec:Schedulers}
Each client has a scheduler that assigns requests to execute at each step.

\subsubsection{LLM Scheduler}\label{subsubsec:llm scheduler}
Since LLM inference requires multiple steps to complete a request, it requires a special scheduler. The LLM Scheduler comprises three modular components:

\textbf{Batch Manager}: Enforces the batching policy for each forward pass.
\tool currently supports five batching strategies:
\textit{Static Batching} (FasterTransformers~\cite{GitHubNV14:online}),
\textit{Continuous Batching} (Orca/vLLM~\cite{yu2022orca, pagedAttention}),
\textit{Chunked Batching} (Sarathi-Serve, FastGen~\cite{agrawal2024taming, holmes2024deepspeed}),
\textit{Mixed Batching} (Splitwise Prefill~\cite{patel2023splitwise}),
and \textit{Disaggregated Batching} (Splitwise/DistServe~\cite{patel2023splitwise, zhong2024distserve}).
The scheduler also enforces user-defined constraints such as the maximum number of batched tokens or batch size.

\textbf{Request Queuer}: Orders incoming requests according to a packing policy.
\tool supports \textit{First-Come-First-Serve (FCFS)} and \textit{Least Work Left}.

\textbf{KV Manager}: Manages KV memory blocks of ongoing requests, and is responsible for prefix caching and preempting KV cache to external KV storage.

Each component is modular, and users can experiment with custom batching, queuing, or scheduling strategies with minimal effort.

\subsubsection{RAG Scheduler}\label{subsubsec:rag_scheduler} 
For each RAG request, the scheduler needs to take two steps: convert the prompt to an embedding using a DPR and retrieve relevant context from the Vector DB. 
If both steps are performed on the same HW (e.g., CPU), we complete both steps from the given batch before starting another batch.
If there is a dedicated accelerator in the HW cluster (e.g., CPU+GPU), we use dual batching: one batch runs DPR on GPU, and the second batch runs DB lookups on CPU. 
%

\subsubsection{KV Scheduler}\label{subsubsec:kv_scheduler}
The KV Manager for the KV Scheduler maintains a record of all KV cache blocks and their associated metadata, including historical accesses and access permissions. Since KV reads are memory-bound operations, we do not batch them by default. The order of KV reads is decided by the request queue. Users can enable batching for client supporting parallel KV reads.

%% file: algorithms/coordinator_sim.tex

\begin{algorithm}[!t]
\small
\caption{Coordinator Simulation Algorithm}
\label{algo:coordinator_sim}
\begin{algorithmic}[1]
    \STATE \textbf{Initialize:} Client interconnect topology
    \STATE \textit{\textbf{Enqueue} arrival of all requests (\textsc{Stage-Push})}
    \WHILE{\texttt{request\_serviced} $<$ \texttt{request\_accepted}}
        \STATE \textbf{Execute next} discrete event in queue

        \IF{Event is \textsc{Stage-Push}}
            \STATE \textbf{Dispatch} stage to client
            \IF{\textit{Client not allotted}}
                \STATE $\text{Client}_{next} \gets \text{Router(Request)}$
            \ENDIF
            \STATE $\text{Client}_{next}.\texttt{add}(\text{Request})$
            \STATE \textit{\textbf{Enqueue} Client to activate next step if idle (\textsc{Client-Step})}

        \ELSIF{Event is \textsc{Client-Step}}
            \STATE \textbf{Process} client step and completed requests
            \STATE $\texttt{Finished\_Requests} \gets \text{Client.next\_step}()$
            \IF{\texttt{Client} has requests to process}
                \STATE \textit{\textbf{Enqueue} Client for next step (\textsc{Client-Step})}
            \ENDIF
            \FOR{each \texttt{request} finished current \texttt{stage}}
                \IF{\texttt{request} is complete}
                    \STATE \textit{Mark request as serviced}
                \ELSE
                    \STATE $\text{Client}_{next} \gets \text{Router(Request)}$
                    \STATE \textit{Start client-transfer event}
                    \STATE \textit{\textbf{Enqueue} request for next stage (\textsc{Stage-Push})}
                \ENDIF
            \ENDFOR
        \ENDIF

    \ENDWHILE
\end{algorithmic}
\end{algorithm}

%% file: sections/3_4_hw-layer.tex
\vspace{-1em}
\subsection{HW Layer: Cluster Modeling}
\label{subsec:cluster_modeling}
For each client in \tool, we have a corresponding HW cluster. These clusters are logical groups of HW nodes, and we simulate these through one of the four methods: 
\squishlist
\item \textbf{Real Execution Runtime}: Execute the actual pipeline stage/step on real hardware. E.g., DecodeBatch[Req1, Req2]. We maintain a local database of execution times. If the \textit{same} batch's runtime is needed, we can look it up instead of running again\footnote{This is very specific to the LLM inference stack, i.e., runtime from vLLM vs SGLang would require unique databases.}. 
\item \textbf{Empirical Runtime}: Profile various combinations of requests on real hardware and create a database of runtimes. Train an ML prediction model to create a real HW runtime database. The key difference from the previous method is that one does not need access to real HW at all times.
\item \textbf{Analytical Modeling}: We can use empirical runtime-inspired analytical models to model runtime for components with a smaller contribution to overall runtime.
\item \textbf{External Simulator}: We can plug in an external hardware simulator~\cite {xu2026aiconfiguratorlightningfastconfigurationoptimization, bambhaniya2024demystifying} to get the hardware runtime of unavailable systems.
\squishend

\autoref{tab:cluster_impl} shows the methods supported for different hardware clusters. We detail the implementation of each hardware cluster in \autoref{subsec:HW_cluster_impl}

\begin{table}[]
\caption{Hardware cluster sim currently supported in \tool}
\label{tab:cluster_impl}
\resizebox{\columnwidth}{!}{%
\begin{tabular}{|c|c|c|c|c|}
\hline
HW Cluster & Real Execution & Empirical Runtime & Analytical Model & External Simulator \\ \hline
LLM Client& \checkmark & \checkmark & \checkmark & \checkmark \\ \hline
RAG Client& \checkmark & \checkmark & \ding{55} & \ding{55} \\ \hline
KV Client & \checkmark  & \checkmark & \ding{55} & \ding{55} \\ \hline
\end{tabular}%
}
\vspace{-1.5em}
\end{table}





%% file: sections/3_5_outputs.tex
\subsection{Output Metrics}
\label{subsubsec:output_metrics}
\tool collects detailed metrics during simulation to analyze how requests are processed across the system. These metrics inform performance insights and guide system design decisions. We categorize the collected data as follows:

\textbf{Individual Request Metrics:} For every request, we record fine-grained statistics, including:
associated stage metrics (client assignment time, stage start time, stage end time), for prefill and decode, we also maintain each token metrics(scheduled time, token start time and token end time). Based on these measurements, we derive the standard inference quality metrics, including TTFT (Time-To-First-Token), TPOT (Time-Per-Output-Token)  to evaluate whether they satisfy the target SLO(Service-Level Objective). 

\textbf{Scheduler-Level Metrics:} These metrics track the request load queued and processed at each simulation step. This includes: Instantaneous and average queue length, variations in arrival volume scheduling rate, step-wise memory load, and finished requests.

\textbf{Client-Level Metrics:} Each client instance maintains operational statistics through its scheduler. Tracked metrics include: Load and queue size at specific timepoints, Request service rate over time, Estimated power consumption.


\textbf{Coordinator Metrics:} To capture holistic system behavior, we log aggregate statistics collected from individual requests such as, serviced requests information, latency breakdowns (mean, P50, P90, P99), and communication metrics.

These global insights enable comprehensive evaluation of system performance and comparative analysis across configurations and scheduling strategies.

\textbf{Request Tracing and Visualization:} All request-level execution details are encoded in JSON format, capturing each processed stage. This format enables seamless integration with visualization tools, such as Chrome Tracing.

Together, the input datasets and output metrics constitute the foundation of our request modeling pipeline, enabling rigorous, end-to-end analysis of inference system behavior under a wide range of workload conditions.

\vspace{-2mm}
\subsection{Extending \tool.} 
\tool has a highly modular and hierarchical design (illustrated in \autoref{fig:overview_figure}(b)).
This 
allows for the decoupling of AI workload, system \& software layer and hardware components, enabling seamless integration across the stack.
\tool employs base-class abstractions for pipeline stages and hardware clients, ensuring extensibility as workloads evolve. Adding a new stage only requires specifying its parameters and a latency model (\autoref{subsec:cluster_modeling}).

%% file: sections/4_0_implementation.tex
\vspace{-2mm}
\section{\tool Implementation and Validation}
\vspace{-1mm}




\input{sections/4_1_cluster-modeling}
\input{sections/4_2_validation}

%% file: sections/4_1_cluster-modeling.tex

\subsection{Implementation of HW Layer}
\label{subsec:HW_cluster_impl}
For the purposes of our case studies (\autoref{sec:eval}), we developed and integrated the following models for the hardware clusters (shown in \autoref{tab:cluster_impl}).

\subsubsection{LLM Client HW Executor}
\label{subsubsec:LLMclust_modelling}


To model the prefill and decode runtime, we use \textit{empirical runtime} data from hardware clusters to predict their runtimes. We collect over 200k datapoints on running vLLM varying input size, batch size, chunk size (for chunked batching), and tensor parallelism (TP1/TP2/TP4/TP8). 
We create estimators for each model, hardware, and stage using an ensemble of regressors, each trained on a distinct, pre-specified subset of the data rather than stochastic samples. 
Using this method, we obtain an average error of 2.5\% with a median error of less than 1\%. 
This approach is $20$–$50\times$ faster compared to real execution and is much more cost-effective.
%
 %
 
Additionally, \tool also supports \textit{external simulators} (specifically AIConfigurator~\cite{xu2026aiconfiguratorlightningfastconfigurationoptimization} and GenZ~\cite{bambhaniya2024demystifying}) to model hypothetical HW configurations (e.g., etched~\cite{Etched}) for the HW Executor of the LLM Client. 
%
\tool also provides \textit{real execution runtime} via vLLM model execution if the hardware is available. This can be helpful in testing the potential performance of optimizations before real implementation. 
%

\subsubsection{RAG Client HW cluster} RAG runtime 
consists of i) Converting the input query into a search space embedding, ii) Re-ranking the top k documents, and iii) Retrieving documents. 
%
%
%
We use the LangChain implementation to collect \textit{empirical runtime} data for running RAG. 
\footnote{The RAG retrieval latency is dependent on the number of documents, the retrieval algorithm. For this work we choose the default configuration of langchain running on 64 core x86 CPU.}

\subsubsection{KV Retrieval HW Cluster} \autoref{subsubsec:kv_scheduler} defines the order of KV retrieval. \tool models the KV cache retrieval cluster with a multi-level memory hierarchy (e.g., GPU HBM $\mapsto$ CPU $\mapsto$ SSD). We have collected the runtime data for KV retrieval at various levels depending on the KV block organization and the number of tokens retrieved. We train a linear regression model on this retrieval latency to predict the runtime. 

\subsection{Accelerated Large-scale Simulation}
We built \tool from the ground up to support large-scale design space exploration across thousands of configurations for streaming, end-to-end inference workloads deployed in the real world. To make the such studies practical and scalable, simulation efficiency is a key design goal, as wall-clock runtime grows with both the number of clients and the number of simulated steps.

Simulation runtime further depends on the cluster runtime backend. Using \toolns's default ML-based models is relatively fast, while selecting a more sophisticated backend, such as a cycle-accurate simulator, incur much higher runtime in exchange of greater fidelity.

To accelerate simulator execution, we use a combination of hash maps and SQL databases to cache the latency of each step in the LLM cluster. This is particularly effective for requests with long decode lengths, where consecutive steps often exhibit similar latencies. Users can also trade off simulation fidelity and runtime by controlling how many consecutive steps reuse cached latency values.

Overall, \tool enables rapid exploration of the design space across diverse hardware and software configurations for their target workloads, without incurring prohibitive GPU costs.

%% file: sections/4_2_validation.tex
\subsection{\tool Validation}
\label{subsec:hermes_validation}


Next, we demonstrate \toolns’s fidelity with different cases.
We perform end-to-end validations ( Aggregated, Disaggregated) of \tool on various platforms and models. We also validate the implementation of LLM Cluster and KV retrieval on different hardware. 

%
 

\input{figures/tex/vllm_mist_cdf_validation.tex}


\subsubsection{Aggregated  Validation} We evaluate \tool using an online workload that has a Poisson distribution at the arrival rate of 5 requests per second (RPS).
Using vLLM, we profile Llama3-70B on \textbf{H100x8}, Qwen3-32B on \textbf{L40Sx2} and Llama3-8B on \textbf{TPUv6e}, all running the ShareGPT~\cite{ShareGPT} dataset.
\autoref{fig:mist_online_serving_validation} shows that the end-to-end runtime of 100 requests.\tool achieves high fidelity, with less than 2\% average error across different combinations of HW and models. 
\toolns's accuracy is significantly better than Vidur\footnote{Vidur doesn't support Google TPUs.} and LLMServingSim2.0 because they profile only the major kernels individually. In real deployments, vLLM Serving Engine uses pre-compiled graphs~\cite{vLLMCUDAGraphs} to reduce LLM inference forward pass time. This can't be captured when reconstructing the latency from individual operators.
Additionally, since both Vidur and LLMServingSim2.0 use operator-level ML predictors, the error is aggregated across layers and tokens, hence impacting end-to-end latency.

\autoref{fig:mist_online_serving_validation}.d shows the simulation time for each LLM inference simulator running the three benchmarks. Vidur requires a significant amount of time—about 400 seconds for pre-training before each run. \tool achieves the fastest simulation runtime while maintaining fidelity due to i.) smarter caching of HW cluster runtimes, ii.) avoiding repeated loading of individual kernels by capturing inference forward-pass latency at the granularity of precompiled graphs.

\subsubsection{Disaggregated Validation} To validate the effectiveness of disaggregated client serving, we utilize real request traces collected from the Azure platform ~\cite{azurellmtrace2023}. As a reference, we compare against Splitwise~\cite{vllm_splitwise2024}, which implements disaggregated serving on top of vLLM for small-scale evaluation. While direct experiments on 80- and 160-GPU clusters are not available to us, splitwise-sim extrapolates from their vLLM prototype to evaluate the same architecture at larger scales. We therefore treat it as a valid simulation-based baseline.

\begin{figure}[!bhtp]
    \centering    \includegraphics[width=1\linewidth]{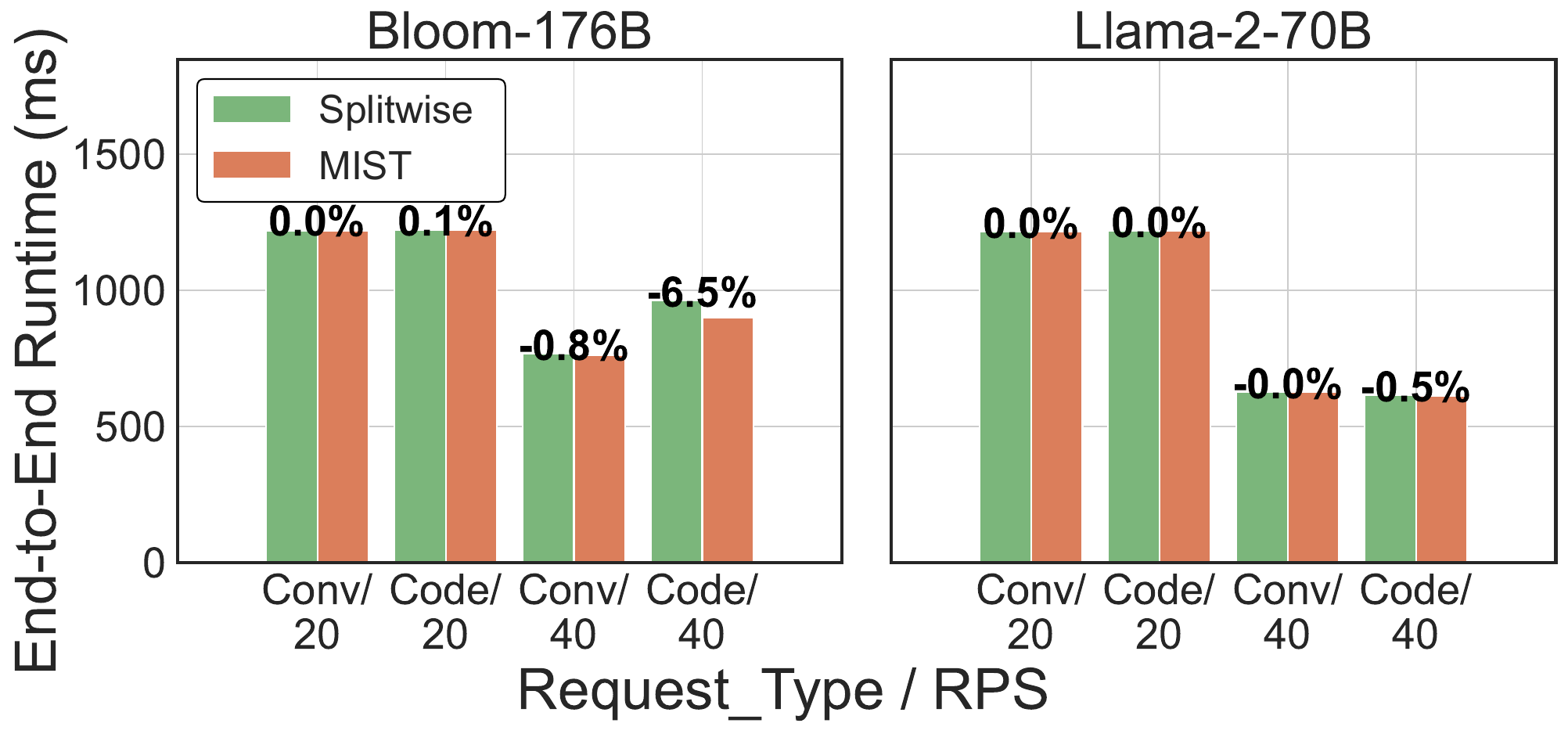}
    \caption{End-to-end validation results comparing Splitwise and \tool on an 80-GPU system configured with 8TP.}
    \label{fig:mist_vs_splitwise validation}
\end{figure}

We simulate two different models(Llama-2-70B and Bloom-176B) on an \textbf{80xH100} node configuration with 8 prefill clients and 2 decode clients under different request distributions (RPS=20 and RPS=40). Across use cases, we observe minor differences in modeling ($<$ 6\% maximum error) as shown in \autoref{fig:mist_vs_splitwise validation}. This difference in runtime arises from communication, as splitwise-sim employs a dummy link-based communication model with a specified lower-bound bandwidth. In contrast, we use Astra-sim to model client communication, which introduces slight differences in overall runtime. \footnote{It should be noted that for this validation, we used splitwise-sim's LLM runtime predictor, as our aim is to show the correctness of request orchestration by the global coordinator.}

\input{figures/tex/runtime_comparison_var_input_len}
\subsubsection{LLM Cluster Modeling Validation}
To evaluate the accuracy of our ML-Based LLM Cluster Modeling (\autoref{subsubsec:LLMclust_modelling}), we compare the \tool per step runtime prediction and Vidur's predicted runtime against the ground truth value from the vLLM running of Llama2-70B on \textbf{H100x8}.
%
\autoref{fig:runtime_comparison_var_input_len} compares the error of \tool and Vidur. \tool's ML-Assisted LLM cluster modeling accuracy is significantly better than  Vidur's runtime prediction. Upon further inspection, we identified two main causes of error in Vidur: (i) using operator-level ML predictors results in error accumulating across layers and operators. (ii) ignoring the kernel launch overheads and smaller operations. \tool uses a predictor trained on the LLM engine's model execution time, which accounts for all the CPU and kernel launch overheads and avoids error cascading.

\subsubsection{KV Retrieval Cluster Modeling Validation}
To evaluate the accuracy of our KV retrieval cluster in \toolns, we compare the measured and \tool predicted latencies for both NVMe SSDs and DDR4 main memory. Sequential reads were performed using the \texttt{fio} benchmarking tool \cite{fiobenchmark_google} across a range of block sizes, from 256 KB to 1GB. Average read latency was recorded for each size.
%
%
%
As shown in \autoref{fig:retreival_latency}, \tool predicts the trend of measured latency closely across both memory tiers. 

%% file: figures/tex/vllm_mist_cdf_validation.tex
\begin{figure}[!t]
    \centering
    \includegraphics[width=1\linewidth]{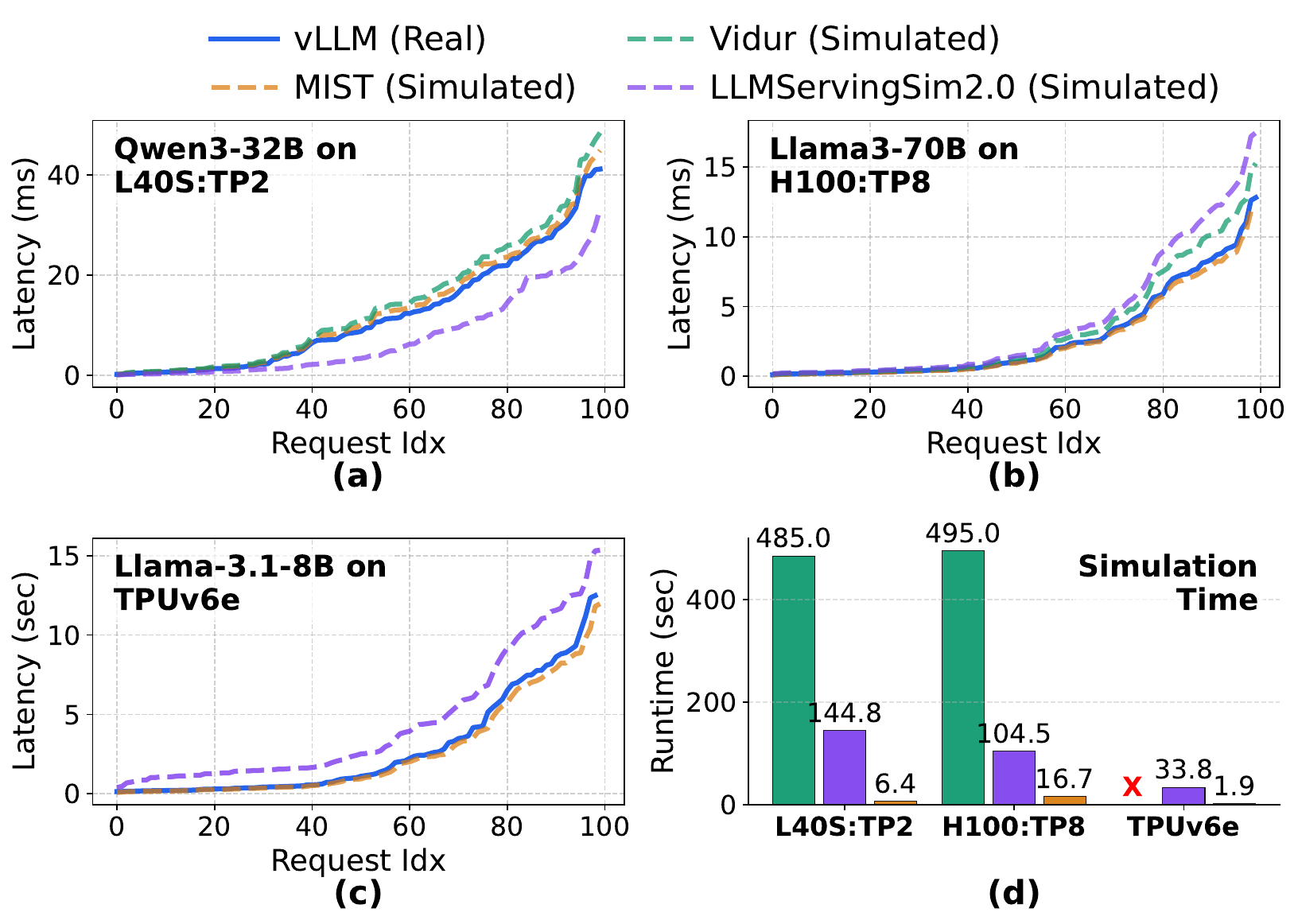}
    \caption{(a-c) Comparing the end-to-end latency on real HW against LLM inference simulators. (d) Simulation time of different LLM inference simulators.
    }
    \vspace{-2em}
\label{fig:mist_online_serving_validation}
\end{figure}


%% file: figures/tex/runtime_comparison_var_input_len.tex
\begin{figure}[!t]
    \centering

    \begin{subfigure}[t]{0.48\columnwidth}
        \centering
        \includegraphics[width=\linewidth]{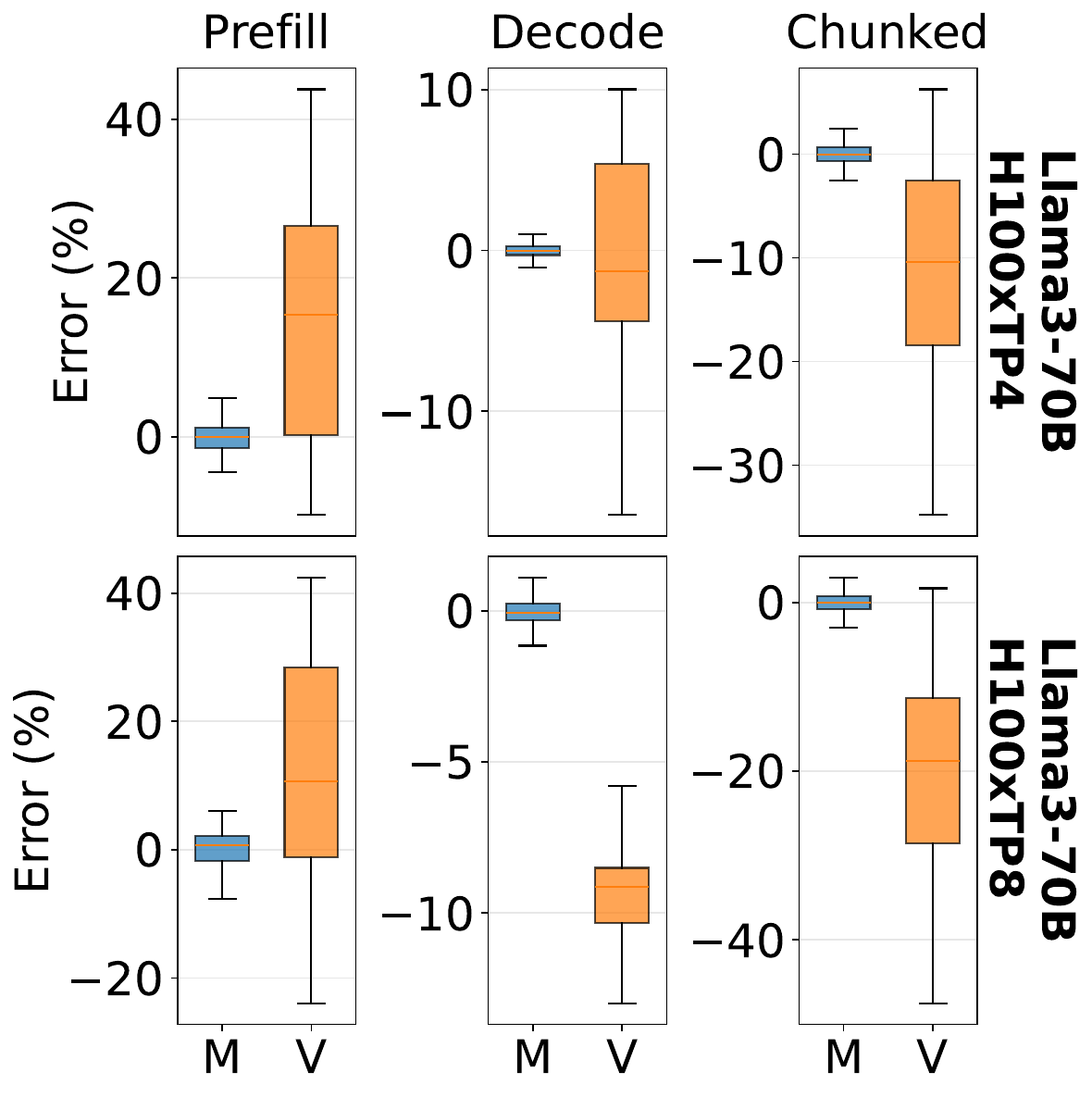}
        \caption{Comparison of \tool:{M} predicted and Vidur:\textbf{V} predicted runtime against vLLM runtime for 1k requests generating 150k tokens.} 
        \label{fig:runtime_comparison_var_input_len}
    \end{subfigure}
    \hfill
    \begin{subfigure}[t]{0.48\columnwidth}
        \centering
        \includegraphics[width=\linewidth]{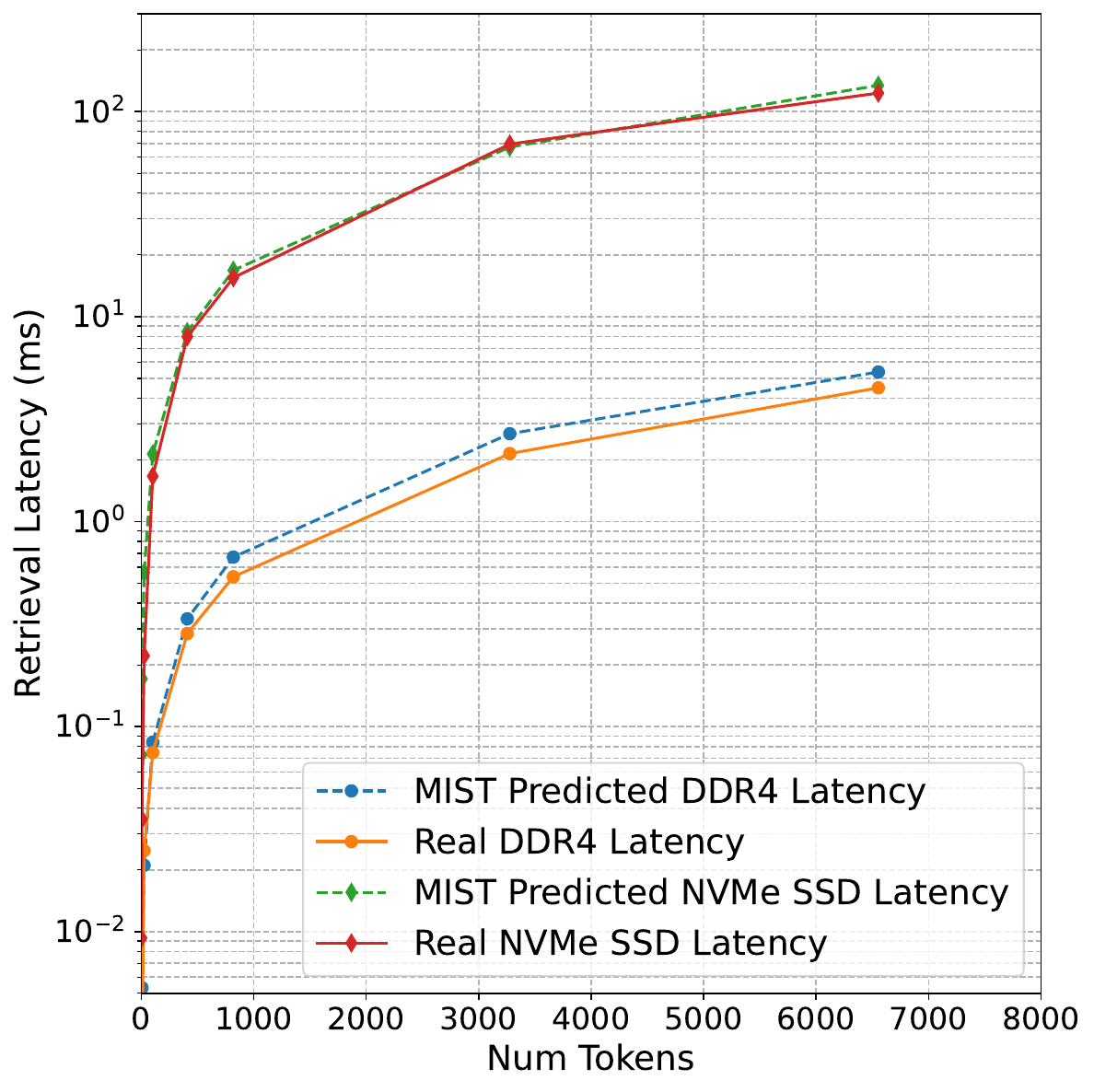}
        \caption{Validating KV Retrieval cluster against real memory devices. We use Llama3-70B KV cache retrieval as an example.}
        \label{fig:retreival_latency}
    \end{subfigure}

    \caption{Hardware Cluster Validation}
    
    \vspace{-1.5em}
\end{figure}

%% file: sections/5_0_eval_isca2026.tex
\vspace{-1em}
\section{Evaluation}
\label{sec:eval}

We demonstrate the value of \tool across two use cases: (i) Optimizing LLM Inference with heterogeneous hardware and (ii) KV cache storage hardware design exploration.

\input{sections/5_1_mist_heterogenous_dse}


\input{sections/5_2_Remote_KV_cache_Storage.tex}

%% file: sections/5_1_mist_heterogenous_dse.tex
\subsection{\tool for optimizing LLM Inference with heterogeneous hardware}

\label{sec:llm_engine_optimization}

\input{tables/deployment_usecase.tex}

In this section, we demonstrate how \tool can efficiently guide the deployment of multi-stage LLM use cases with diverse token distributions across off-the-shelf heterogeneous hardware.
Modern LLM serving engines~\cite{kwon2023efficient,zheng2024sglang,nv_dynamo} expose a wide variety of configurable parameters (tuning knobs) that significantly impact deployment efficiency.
Using \toolns, we systematically explore combinations of these knobs to maximize generated tokens per second per dollar, for requests that meet their latency SLOs.

\textbf{Search Space:} We search the deployment space across the following configurable dimensions:
(i)~\textbf{hardware SKUs} spanning Nvidia (H200-SXM, B200-SXM, GB300), AMD (MI350X, MI355X), Google (TPUv6e, TPUv7), and Etched (Soho)\footnote{We assume Etched Soho has similar memory and network characteristics as Nvidia GB300, with 5$\times$ the compute units~\cite{etched_analysis}.};
(ii)~\textbf{model parallelism} with tensor parallelism TP~$\in~\{1,2\}$;
(iii)~\textbf{batching strategies} including chunked and disaggregated prefill; and
(iv)~\textbf{prefill-to-decode client ratios} for disaggregated deployments.
We use cloud rental pricing\footnote{We use pricing from CoreWeave for Nvidia GPUs, TensorWave for AMD GPUs, and Google Cloud for TPUs. We estimate the price for publicly unavailable HW based on similar-sized, spec-ed chips. } (per accelerator per hour) to estimate search and deployment costs: H200-SXM \$6.31, B200-SXM \$8.60, GB300 \$12.00, Etched Sohu \$12.00, MI350X \$7.50, MI355X \$9.50, TPUv6e \$3.00, and TPUv7 \$8.00. We use Nvidia's AIConfigurator's roofline estimation~\cite{xu2026aiconfiguratorlightningfastconfigurationoptimization} as the LLM Cluster runtime simulator, ensuring all HW SKUs are modeled with the same backend and based on their raw capabilities. 

We evaluate \tool across two representative multi-stage use cases (\autoref{tab:deployment_usecase}).
\autoref{fig:optimization_comparison} compares the best output generation throughput/\$~and deployment cost of the top configuration across vendors for both use cases.

\subsubsection{Chat Conversation}
\label{subsec:long_context}

\textit{Use case: Regular chat-based conversation with user prompts.}
We use the Narrative QA dataset~\cite{kocisky-etal-2018-narrativeqa} as a representative workload, where queries pose questions about movies and books to elicit free-form answers.

\textit{Experimental Setup:}
We use Qwen3-32B~\cite{qwen3technicalreport} and search over deployment configurations (up to 8 Nodes) at a query injection rate of 100~RPS.
Given the streaming chat generation use case, we enforce a TTFT SLO of P99~$\leq$~250~ms.

Our search yields 1{,}567 valid configurations.
\autoref{fig:chatbot_search} shows the token generation throughput across all explored deployments.
Among single-vendor configurations, TPUv7~(2P:6D) and B200-SXM~(2P:6D) achieve the highest raw throughput owing to their superior TFLOPS and memory bandwidth.
However, optimizing for generated tokens per second per dollar favors a heterogeneous, multi-vendor mix: 2$\times$MI350X (prefill) paired with 6$\times$H200-SXM (decode).
This configuration produces 22\% more tokens/\$ at 17.4\% lower per-hour deployment cost than a single-vendor configuration, demonstrating that cross-vendor hardware pairings can unlock cost-efficiency gains unavailable within any single vendor's product line.

\input{figures/tex/heterogenous_dse_search}

\input{figures/tex/optimization_comparision}

\subsubsection{Code Generation with High Prefix-KV Reuse}
\label{subsec:prefix_kv}

\textit{Use case:} LLM-assisted code generation has become a common practice in modern software engineering.
When generating code with multi-file dependencies, the relevant files must be supplied as context for the prompt.
The KV cache for most files can be pre-computed offline and reused across requests via prefix caching~\cite{moduler_prefix_caching}.
We use the GitHub Code dataset~\cite{hf_github-code-2025} to simulate requests that edit small files using pre-calculated KV context from other files in the same repository.

\textit{Experimental Setup:}
We use Qwen3-32B~\cite{qwen3technicalreport} and search over deployment configurations (up to 8 GPUs) at 100~RPS.
Given the large historical context window, we enforce a TTFT SLO of P99~$\leq$~1.5~s.

The search yields 1{,}878 valid configurations (\autoref{fig:code_gen_search}).
While most requests benefit from a large fraction of pre-computed KV context, roughly 10\% of requests still require full KV recomputation over long context sequences---so-called \textit{whale} requests---making high prefill compute capacity critical.
Among single-vendor deployments, the Etched Soho chip achieves the highest raw throughput owing to its exceptional compute density.
However, optimizing for generated tokens/\$~reveals a more nuanced picture: a heterogeneous Etched~+~TPUv6e configuration (1 prefill client~$\times$~TP2 + 3 decode clients~$\times$~TP2) delivers the best cost-efficiency.
The intuition is clear - Etched Soho absorbs the compute-intensive KV recomputation for whale requests, while TPUv6e, which offers the highest memory bandwidth per dollar among the candidates, handles the decode stage efficiently.
Optimizing for generated tokens per second per dollar, this Etched-TPUv6e pairing outperforms the best single-vendor alternative, TPUv7 with chunked batching (4 clients, TP2), generating \textbf{49.3\% more tokens/\$} at \textbf{34.4\% lower deployment cost}.



%% file: tables/deployment_usecase.tex
\definecolor{stagePrefill}{RGB}{148,29,42}        
\definecolor{stageDecode}{RGB}{198,42,114}      
\definecolor{stageRAG}{RGB}{213,232,212}           
\definecolor{stageReason}{RGB}{255,230,204}        
\definecolor{stageKV}{RGB}{150,86,53}           

\newcommand{\stagebox}[2]{%
  \tikz[baseline=(s.base)]
    \node[
      fill=#1, 
      rounded corners=2pt,
      inner sep=2.2pt
    ](s){\textcolor{white}{#2}};
}

\begin{table*}[]
\centering
\caption{LLM use cases with different multi-stage pipelines.}
\label{tab:deployment_usecase}

\resizebox{\linewidth}{!}{
\begin{tabular}{|c|c|c|c|ccc|ccc|}
\hline
\multirow{2}{*}{Usecase} &
\multirow{2}{*}{Datasets} &
\multirow{2}{*}{Stages} &
\multirow{2}{*}{\begin{tabular}[c]{@{}c@{}}\\ \# Queries\end{tabular}} &
\multicolumn{3}{c|}{\# Prefill Tokens} &
\multicolumn{3}{c|}{\# Decode Tokens} \\ \cline{5-10}
 &  &  & 
 & \multicolumn{1}{c|}{Mean} 
 & \multicolumn{1}{c|}{Median} 
 & P95
 & \multicolumn{1}{c|}{Mean} 
 & \multicolumn{1}{c|}{Median} 
 & P95 \\ \hline


Chat Conversation &
Narrative QA~\cite{kocisky-etal-2018-narrativeqa} &
\stagebox{stagePrefill}{Prefill}\stagebox{stageDecode}{Decode} &
288 & 252 & 484 & 1597 & 218 & 232 & 332\\ \cline{1-4}
Code Generation &
Github-Code~\cite{hf_github-code-2025} &
\stagebox{stageKV}{KV Retrieval}\stagebox{stagePrefill}{Prefill}\stagebox{stageDecode}{Decode} &
465k & 26979 & 16204 & 83920 & 1088 & 329 & 2753 \\
 \hline
\end{tabular}}
\vspace{-1.5em}
\end{table*}

%% file: figures/tex/heterogenous_dse_search.tex
\begin{figure}[t]
    \centering
    \begin{subfigure}{0.9\linewidth}
        \centering
        \includegraphics[width=\linewidth]{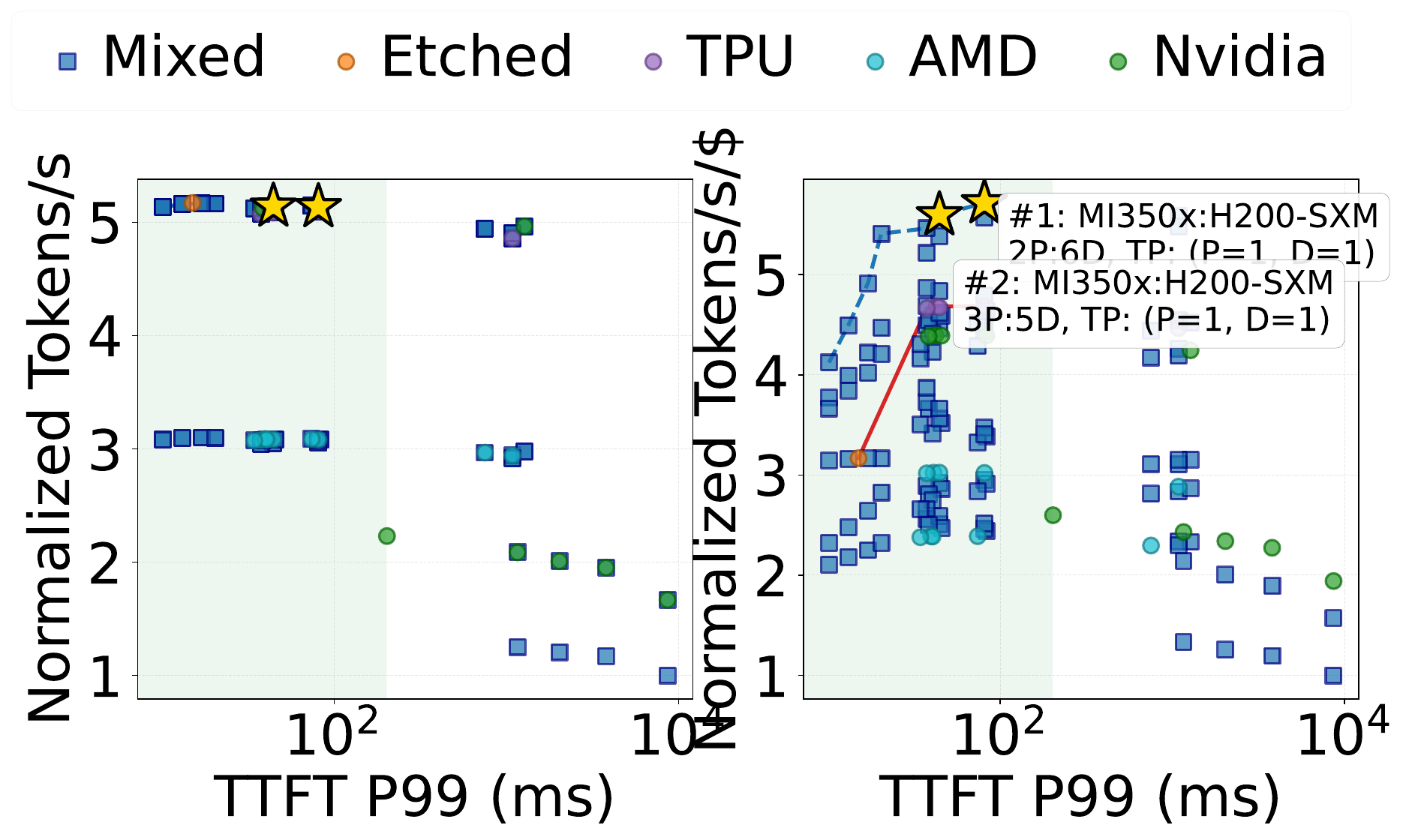}
        \caption{Chat Conversation.}
        \label{fig:chatbot_search}
    \end{subfigure}
    \begin{subfigure}{0.9\linewidth}
        \centering
        \includegraphics[width=\linewidth, trim=0 0 0 80pt, clip]{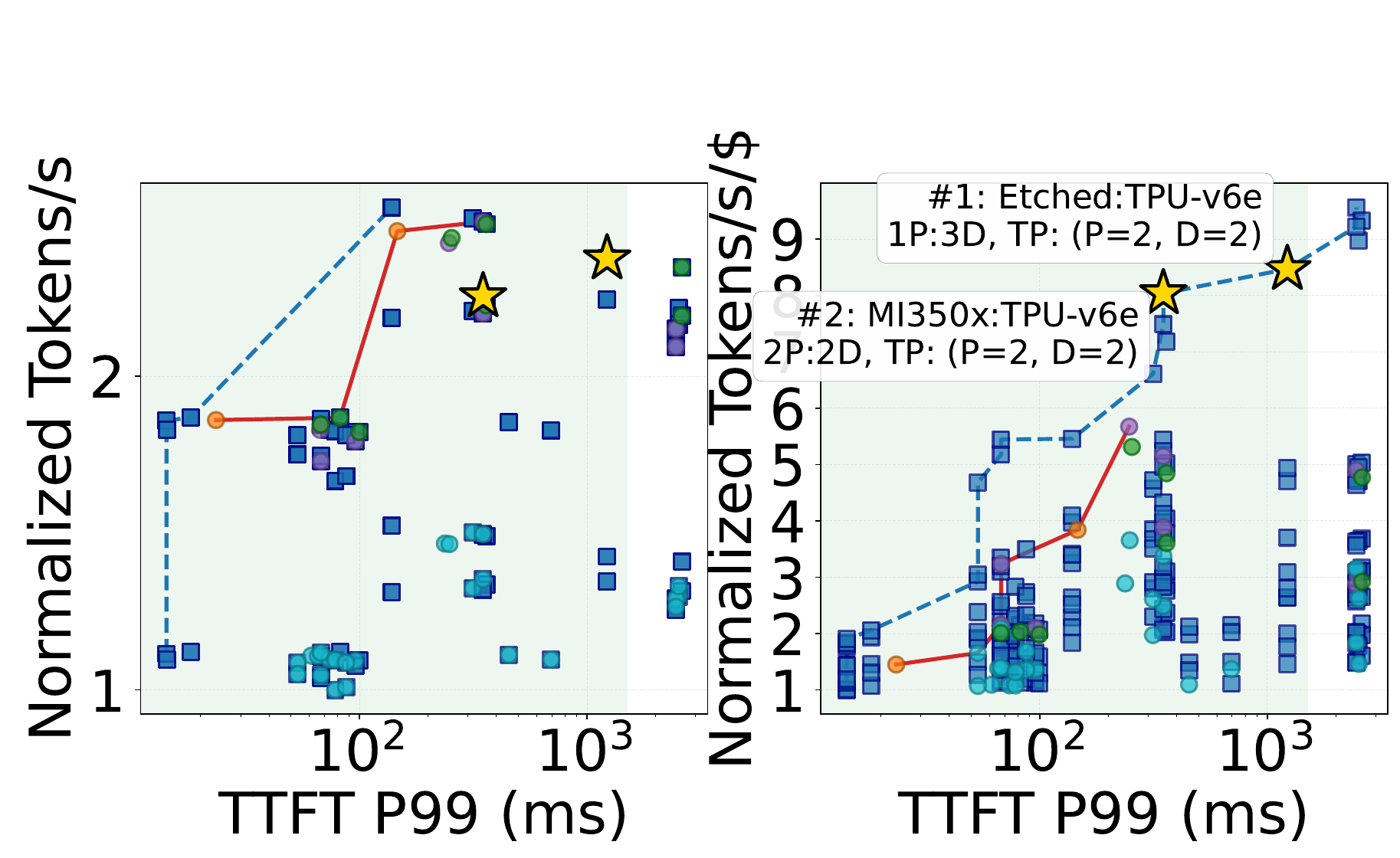}
        \caption{Code generation with High Prefix-KV Reuse.}
        \label{fig:code_gen_search}
    \end{subfigure}
    \caption{Search-space exploration for different use cases.
    Output generation throughput is normalized against the lowest throughput configuration. 
    Stars represent the top 2 optimal configurations found by \toolns.}
    \vspace{-1.5em}
\end{figure}

%% file: figures/tex/optimization_comparision.tex
\begin{figure}
    \centering
    \includegraphics[width=1\columnwidth]{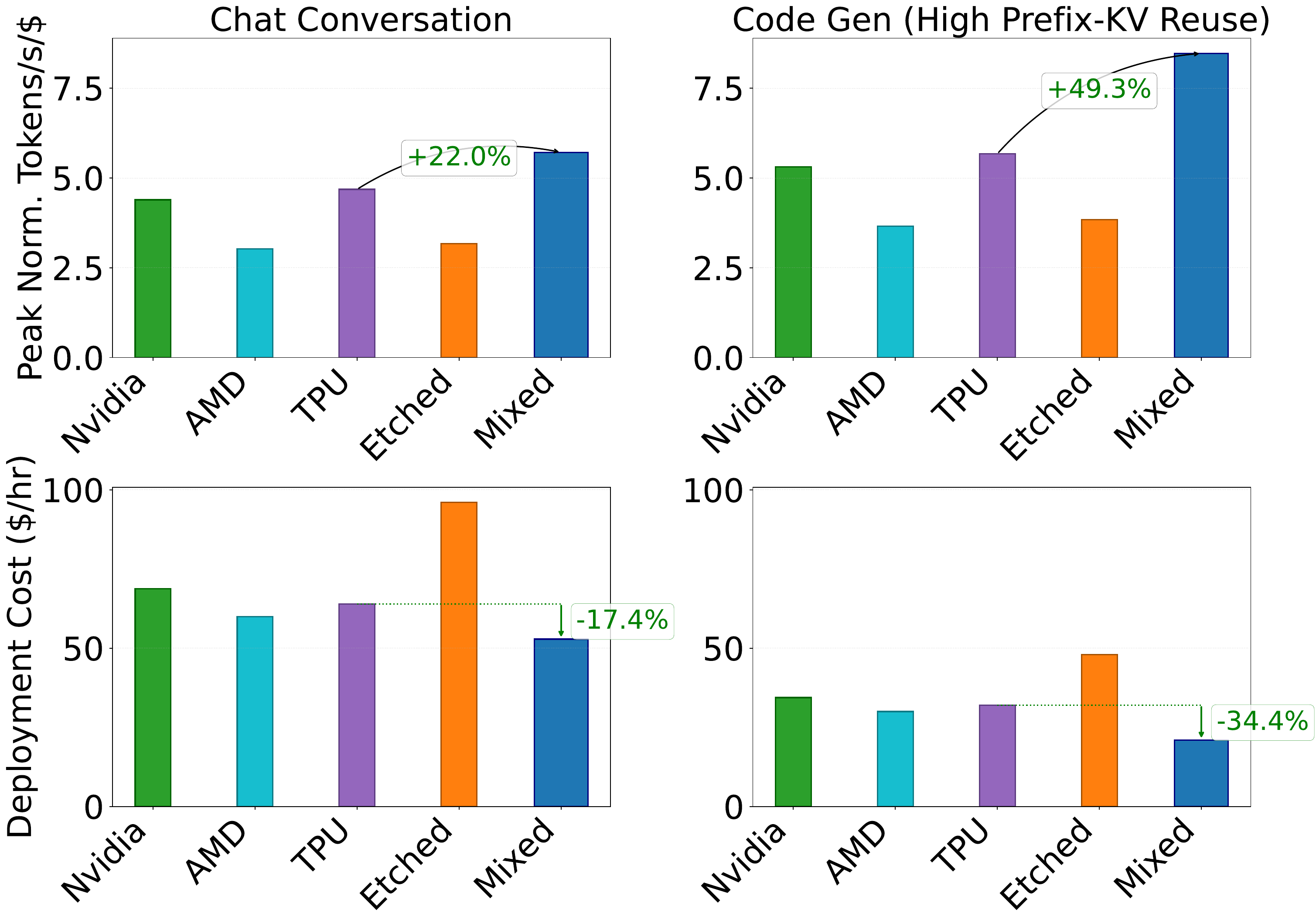}
    \vspace{-1.5em}
    \caption{LLM deployment search results for different use-cases.}
    \label{fig:optimization_comparison}
    \vspace{-1.5em}
\end{figure}

%% file: sections/5_2_Remote_KV_cache_Storage.tex
\vspace{-0.5em}
\subsection{\tool for KV cache storage hardware design exploration.}
\label{subsec:c6_kv_storage}

\input{figures/tex/memory_storage_solution_types}








Efficient KV cache retrieval is crucial for LLM serving, as faster KV retrieval means lower TTFT. 
%

\textbf{Key Question: Where should we store KV Cache?}

\textbf{Target Usecase:} The analysis addresses two principal scenarios: \textit{i) private key-value (KV) caches}—designed for individual user contexts(e.g., ChatGpt). These private KV caches can be accessed by future queries from the same user.
\textit{ii) Shared KV caches}—often used in multi-user settings (e.g., enterprise AI or shared codebases) to enable multiple users to access a large corpus (~$O(10^{10})$ tokens) of documents or code.
Generally, these caches have hotspots that are accessed much more frequently than the rest of the KV cache.

\noindent \textbf{Hardware design space:} 
Current AI serving racks with storage memory use different caching solutions. We classify these into three choices (\autoref{fig:cache_storage_solutions}): (1) a dedicated cache per client, inspired by dedicated DDR on the host of the GPU.  (2) a platform-level shared cache with shared access by 2-8 clients, and (3) a rack-level shared cache with shared access by 32-64 clients, inspired by SSD memory in each rack. These racks are connected by data center network(DCN).
Each configuration offers distinct trade-offs among capacity, bandwidth, and latency.

\noindent \textbf{Experimental Setup:} We evaluate these architecture choices on 256 GPUs (128 clients with H100:TP2) distributed across 4 racks, interconnected via NVLink and PCIe. The experiments simulate short (4K tokens) and long (24K tokens) KV cache retrieval workloads for both private and shared contexts. Requests follow the AzureConv trace~\cite{azurellmtrace2023}.
%
We examine five distinct storage architectures, detailed in \autoref{tab:exp_setup}, ranging from dedicated local memory to a baseline of full recomputation. End-to-end request serving latency serves as the primary evaluation metric to assess system performance across these configurations.

\begin{table}[h]
\centering
\caption{Storage Architecture Configs. Bandwidth (BW) represents the speed of data retrieval or transfer.}
\label{tab:exp_setup}
\resizebox{\columnwidth}{!}{%
\begin{tabular}{@{}clcc@{}}
\toprule
\textbf{Case} & \textbf{Architecture Type} & \textbf{Capacity / BW} & \textbf{Access / Notes} \\ \midrule
A & Dedicated Per-Client (Choice 1) & 1 TB / 128 GB/s & LPDDR-based; Local access \\
B & Platform-Shared (Choice 2) & 4 TB / 32 GB/s & Shared by 4 clients \\
C & Rack-Shared (Choice 3) & 32 TB / 2 GB/s & Shared by 32 clients \\
D & Rack-Shared (Choice 3) + DCN & 32 TB / 2 GB/s & Inter-rack transfer @ 128GB/s \\
E & No Cache & N/A & Full KV recomputation \\ \bottomrule
\end{tabular}%
}
\end{table}

\autoref{fig:memory_retrieval_strategies} shows the end-to-end latency cdf for the design space. 
\textbf{For private KV caches}, we find that the platform-level shared cache (Case B) or dedicated LPDDR (Case A) offers the best P90 request latency for both short and long requests. This is because, under normally distributed requests arrivals, clients' loads remain balanced. In contrast, performance of Cases C, D, E degrades as KV context length increases because of lower bandwidth and higher latency.

\textbf{For shared KV caches}, storing KV cache at the platform level (B) or in host memory (A) is a very poor choice as it leads to severe imbalance in the system; instead, using rack-level SSD-based (Case C) is the best option, as it delivers higher capacity and maintains acceptable despite a modest reduction in per-client bandwidth.
For larger KV caches, storing them at the platform level (B) becomes attractive due to superior memory BW, resulting in lower latency. 
\input{figures/tex/memory_retrieval_case_studies}

\vspace{-0.5em}
  \begin{tcolorbox}[colback=black!10,colframe=black]
   \noindent $\bullet$ Dedicated (A) or Platform-level shared cache (B) is great for private KV caches as it balances speed and resource sharing.
   
   \noindent $\bullet$ Rack-level shared cache (C) is optimal for shared global KV caches: Provides low-latency access and efficient inter-client sharing.
   
\end{tcolorbox}
\vspace{-0.5em}

%% file: figures/tex/memory_storage_solution_types.tex
\begin{figure}[!bthp]
\vspace{-1em}
    \centering
    \includegraphics[width=1\linewidth]{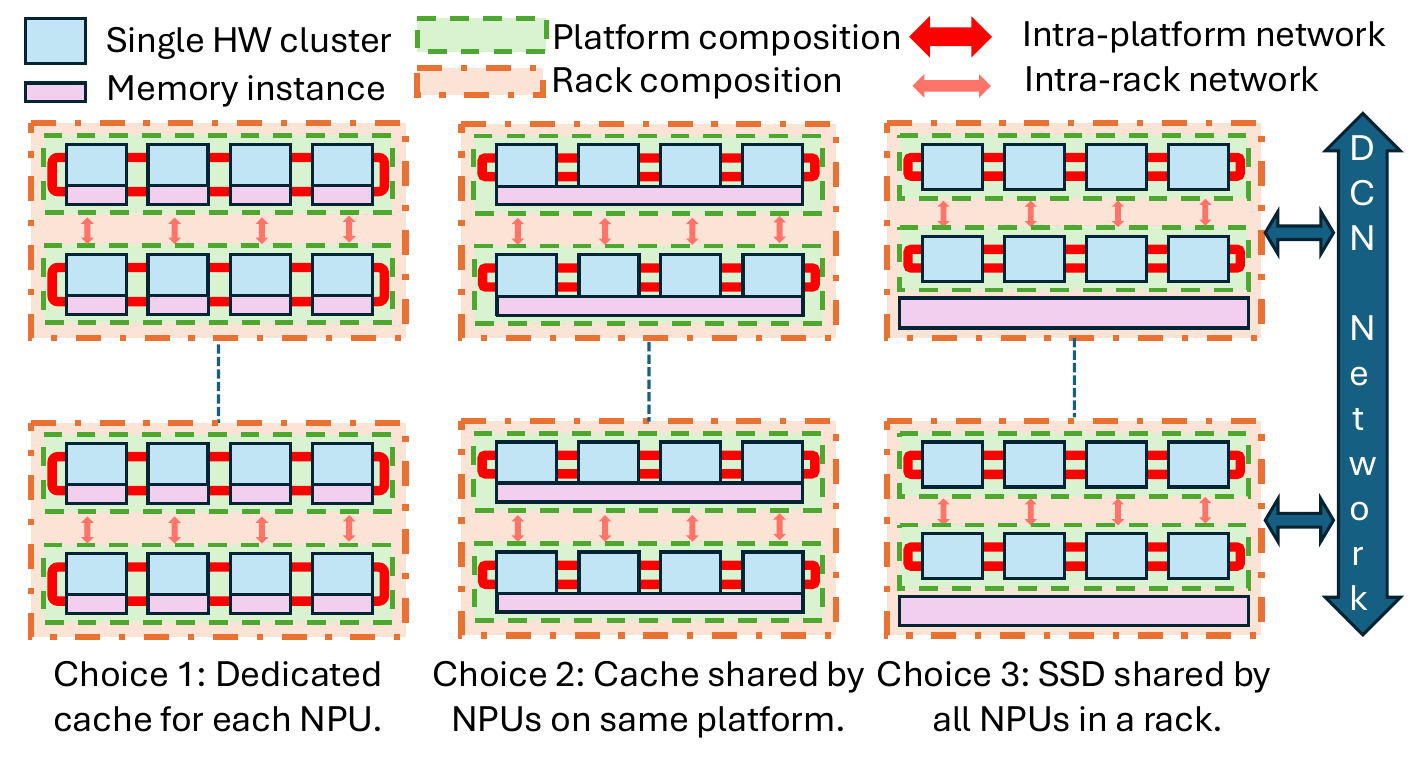}
    \vspace{-1em}
    \caption{\textsc{Design choices for KV cache storage.}}
    \label{fig:cache_storage_solutions}
    \vspace{-1.5em}
\end{figure}

%% file: figures/tex/memory_retrieval_case_studies.tex
\begin{figure}[!t]
\vspace{-1em}
    \centering    \includegraphics[width=1\linewidth]{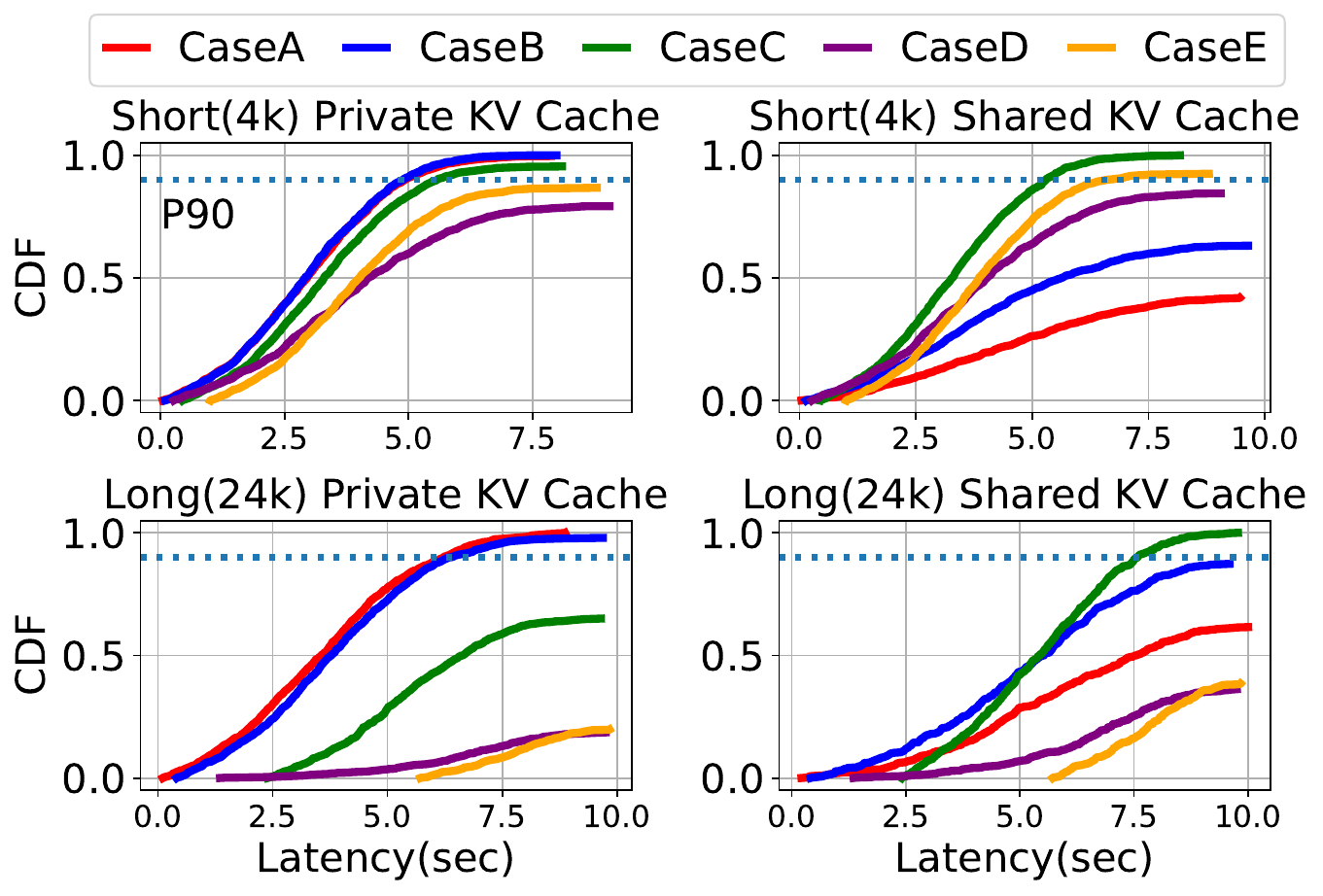}
    \vspace{-1.5em}
    \caption{Comparing different platform architectures for storing past cache storage. Serving 128 clients of Llama-3.1-70B using 4 HGX Racks(64xH100 per Rack).} \label{fig:memory_retrieval_strategies}
    \vspace{-1.5em}
\end{figure}

%% file: sections/6_related_work.tex
\vspace{-1mm}
\section{Related Work}
\label{sec:related}
\vspace{-1mm}
\autoref{tab:related_works} compares various frameworks simulating the LLM inference serving stack. LLMCompass\cite{zhang2023hardware} and GenZ\cite{bambhaniya2024demystifying} provide detailed modeling capabilities with optimizations but are limited to single-client configurations.
Vidur~\cite{agrawal2024vidurlargescalesimulationframework} supports multi-client simulation via discrete-event simulation and ML-based runtime prediction, but relies on analytical roofline models that abstract framework-specific behavior. It assumes homogeneous clients and is restricted to real hardware configurations with aggregated batching strategies only, lacking support for disaggregated prefill--decode hardware.
Splitwise-sim~\cite{splitwisesim2024} models three pools for hardware clients representing prefill, decode and mixed pool. It does not model chunked batching and modeling of end-to-end pipeline stages.
AIConfigurator~\cite{xu2026aiconfiguratorlightningfastconfigurationoptimization} automates configuration selection across multi-framework LLM serving systems, such as vLLM and TensorRT-LLM, using an extensive set of real GPU measurements to predict end-to-end performance. However, it assumes fixed batch sizes and fixed input and output sequence lengths, and therefore cannot evaluate serving configurations under streaming requests from real-world datasets such as the Azure trace~\cite{azurellmtrace2023}. In addition, it models only prefill and decode on independent GPUs, without modeling network routing across heterogeneous systems or supporting multi-stage pipeline simulation.
Notably, \tool integrates AIConfigurator as an \textit{external simulator}, leveraging its extensive measurement dataset collected across diverse hardware platforms.
LLMServingSim~2.0~\cite{cho2026llmservingsim20unifiedsimulator} extends LLMServingSim to support heterogeneous accelerator pools (e.g., TPUs, CIMs), profile-based execution, and more detailed modeling of KV cache movement, prefix reuse, and expert-parallel routing for MoE models. While it substantially improves hardware diversity and routing policies over its predecessor, it remains restricted to single-model deployments and does not model advanced multi-stage pipeline stages.
Overall, Vidur, Splitwise-sim, AIConfigurator, and LLMServingSim~V2 all fall short in modeling advanced, multi-stage LLM inference pipelines.
\textit{In contrast, \tool is the first simulator designed to support end-to-end modeling of real-world LLM inference pipelines across diverse HW configurations, while being significantly faster, offering higher simulation fidelity, and providing a modular architecture that enables flexible composition of pipeline stages and hardware backends.}
\input{tables/related_works_compact}

%% file: tables/related_works_compact.tex
%
%


\begin{table}[t]
\centering
\caption{\textsc{Comparison of LLM Serving Simulators}}
\setlength{\tabcolsep}{3.5pt}
\renewcommand{\arraystretch}{1.1}
\resizebox{\columnwidth}{!}
{\begin{tabular}{l|ccc|cc|ccc}
\hline
\multirow{2}{*}{\textbf{Simulator}} & \multicolumn{3}{c|}{\textbf{Workload}} & \multicolumn{2}{c|}{\textbf{SW}} & \multicolumn{3}{c}{\textbf{HW}} \\ \cline{2-9} 
 & TR & PS & MM & RR & B & HT & MP & TKV \\ \hline
AIConfigurator~\cite{xu2026aiconfiguratorlightningfastconfigurationoptimization} & \xmark & \xmark & \xmark & \xmark & \cmark & \pmark & \pmark & \xmark \\
Vidur~\cite{agrawal2024vidurlargescalesimulationframework} & \cmark & \xmark & \xmark & \cmark & \pmark & \xmark & \xmark & \xmark \\
LLMServingSim~\cite{cho2024llmservingsimhwswcosimulationinfrastructure} & \cmark & \xmark & \xmark & \xmark & \pmark & \pmark & \pmark & \cmark \\
Splitwise-sim~\cite{splitwisesim2024} & \cmark & \xmark & \xmark & \xmark & \pmark & \pmark & \pmark & \xmark \\
LLMServingSim ~2.0~\cite{cho2026llmservingsim20unifiedsimulator} & \cmark & \xmark & \xmark & \cmark & \cmark & \cmark & \pmark & \cmark \\ \hline
\textbf{\tool (ours)} & \cmark & \cmark & \cmark & \cmark & \cmark & \cmark & \cmark & \cmark \\ \hline
\end{tabular}%
}
\begin{flushleft}
{\footnotesize
\textbf{TR}: Trace Replay,
\textbf{PS}: Pipeline stages (end-to-end inference workflow beyond prefill/decode, such as KV Cache Retrieval and RAG),
\textbf{MM}: Multi-Model Serving,
\textbf{RR}: Multi Node Routing, 
\textbf{B}:  Batching (Agg, PD Disagg),
\textbf{HT}: Heterogeneous Systems,
\textbf{MP}: Multiple HW Pools,
\textbf{TKV}: Tiered KV Cache Hierarchy.
\\
\cmark: fully supported,~~\xmark: not supported,~~\pmark: limited or partial support.
}
\end{flushleft}
\vspace{-2.5em}
\label{tab:related_works}
\end{table}

%% file: sections/7_conclusion.tex
\vspace{-3mm}
\section{Conclusion}
\vspace{-1mm}
We present \toolns, an event-driven simulation framework modeling the full LLM serving stack across workload, system-software, and hardware layers, achieving end-to-end fidelity within 6\% of real deployments.
Through case studies, \tool reveals that multi-vendor PD-disaggregated configurations yield up to 49.3\% higher tokens/\$ over homogeneous deployments, and that KV storage hierarchy and eviction policy choices significantly shape tail latency.
%
%
Looking ahead, \tool can guide future hardware roadmap decisions, develop adaptive cross-stage schedulers, and model emerging multi-agent LLM orchestration patterns.




%% file: refs.bib
@misc{gimlet_labs_2026,
author = {Allison Francis},
  title = {Gimlet Labs Targets AI’s Inference Cost Problem},
  url = "https://www.channelinsider.com/ai/gimlet-labs-80m-series-a-ai-inference/",
month = {3},
year = {2026},
  note = "[Online; accessed 2026-04-07]"
}

@misc{vLLMCUDAGraphs,
author = {vllm},
  title = {CUDA Graphs - vLLM},
  url = "https://docs.vllm.ai/en/stable/design/cuda_graphs/",
month = {3},
year = {2026},
  note = "[Online; accessed 2026-04-07]"
}

@misc{nvidia_groq_lpx,
author = {Kyle Aubrey},
  title = {Inside the NVIDIA Vera Rubin Platform: Six New Chips, One AI Supercomputer | NVIDIA Technical Blog},
  url = "https://developer.nvidia.com/blog/inside-the-nvidia-rubin-platform-six-new-chips-one-ai-supercomputer/",
month = {1},
year = {2026},
  note = "[Online; accessed 2026-04-07]"
}

@misc{aws_cerebras_2026,
author = {James Wang},
  title = {Cerebras},
  url = "https://www.cerebras.ai/blog/cerebras-is-coming-to-aws",
month = {3},
year = {2026},
  note = "[Online; accessed 2026-04-07]"
}

@misc{ShareGPT,
  author = {anon823116},
  title = {ShareGPT\_Vicuna Datasets at Hugging Face},
  url = "https://huggingface.co/datasets/anon8231489123/ShareGPT\_Vicuna\_unfiltered",
  year = {2024},
}

@misc{etched_analysis,
author = {Peak FLOPS},
  title = {breaking down Etched's Sohu - by Peak FLOPS - Emilio},
  url = "https://wafer.substack.com/p/breaking-down-etcheds-sohu",
month = {4},
year = {2026},
  note = "[Online; accessed 2026-04-09]"
}

@misc{xu2026aiconfiguratorlightningfastconfigurationoptimization,
      title={AIConfigurator: Lightning-Fast Configuration Optimization for Multi-Framework LLM Serving}, 
      author={Tianhao Xu and Yiming Liu and Xianglong Lu and Yijia Zhao and Xuting Zhou and Aichen Feng and Yiyi Chen and Yi Shen and Qin Zhou and Xumeng Chen and Ilya Sherstyuk and Haorui Li and Rishi Thakkar and Ben Hamm and Yuanzhe Li and Xue Huang and Wenpeng Wu and Anish Shanbhag and Harry Kim and Chuan Chen and Junjie Lai},
      year={2026},
      eprint={2601.06288},
      archivePrefix={arXiv},
      primaryClass={cs.LG},
      url={
https://doi.org/10.48550/arXiv.2601.06288}, 
}

@misc{cho2026llmservingsim20unifiedsimulator,
      title={LLMServingSim 2.0: A Unified Simulator for Heterogeneous and Disaggregated LLM Serving Infrastructure}, 
      author={Jaehong Cho and Hyunmin Choi and Guseul Heo and Jongse Park},
      year={2026},
      eprint={2602.23036},
      archivePrefix={arXiv},
      primaryClass={cs.DC},
      url={
https://doi.org/10.48550/arXiv.2602.23036}, 
}

@misc{Gemini,
      title={Gemini: A Family of Highly Capable Multimodal Models}, 
      author={Gemini Team},
      year={2025},
      eprint={2312.11805},
      archivePrefix={arXiv},
      primaryClass={cs.CL},
      url={
https://doi.org/10.48550/arXiv.2312.11805}, 
}

@misc{zheng2024sglang,
      title={SGLang: Efficient Execution of Structured Language Model Programs}, 
      author={Lianmin Zheng and Liangsheng Yin and Zhiqiang Xie and Chuyue Sun and Jeff Huang and Cody Hao Yu and Shiyi Cao and Christos Kozyrakis and Ion Stoica and Joseph E. Gonzalez and Clark Barrett and Ying Sheng},
      year={2024},
      eprint={2312.07104},
      archivePrefix={arXiv},
      primaryClass={cs.AI},
      url={
https://doi.org/10.48550/arXiv.2312.07104}, 
}

@misc{ahn2024llmmaths,
      title={Large Language Models for Mathematical Reasoning: Progresses and Challenges}, 
      author={Janice Ahn and Rishu Verma and Renze Lou and Di Liu and Rui Zhang and Wenpeng Yin},
      year={2024},
      eprint={2402.00157},
      archivePrefix={arXiv},
      primaryClass={cs.CL},
      url={
https://doi.org/10.48550/arXiv.2402.00157}, 
}

@misc{ChatGPT,
	title        = {ChatGPT},
	author       = {OpenAI},
	journal      = {ChatGPT},
	publisher    = {openAI},
	url          = {https://openai.com/chatgpt}
}

@misc{wei2023cot,
      title={Chain-of-Thought Prompting Elicits Reasoning in Large Language Models}, 
      author={Jason Wei and Xuezhi Wang and Dale Schuurmans and Maarten Bosma and Brian Ichter and Fei Xia and Ed Chi and Quoc Le and Denny Zhou},
      year={2023},
      eprint={2201.11903},
      archivePrefix={arXiv},
      primaryClass={cs.CL},
      url={https://doi.org/10.48550/arXiv.2201.11903}, 
}

@misc{yang2024sweagent,
      title={SWE-agent: Agent-Computer Interfaces Enable Automated Software Engineering}, 
      author={John Yang and Carlos E. Jimenez and Alexander Wettig and Kilian Lieret and Shunyu Yao and Karthik Narasimhan and Ofir Press},
      year={2024},
      eprint={2405.15793},
      archivePrefix={arXiv},
      primaryClass={cs.SE},
      url={https://doi.org/10.48550/arXiv.2405.15793}, 
}

@inproceedings{Lin2025Parrot,
author = {Lin, Chaofan and Han, Zhenhua and Zhang, Chengruidong and Yang, Yuqing and Yang, Fan and Chen, Chen and Qiu, Lili},
title = {Parrot: efficient serving of LLM-based applications with semantic variable},
year = {2024},
isbn = {978-1-939133-40-3},
publisher = {USENIX Association},
address = {USA},
booktitle = {Proceedings of the 18th USENIX Conference on Operating Systems Design and Implementation},
articleno = {50},
numpages = {17},
location = {Santa Clara, CA, USA},
series = {OSDI'24},
url = {https://doi.org/10.48550/arXiv.2405.19888}
}

@misc{agrawal2024vidurlargescalesimulationframework,
	title        = {Vidur: A Large-Scale Simulation Framework For LLM Inference},
	author       = {Amey Agrawal and Nitin Kedia and Jayashree Mohan and Ashish Panwar and Nipun Kwatra and Bhargav Gulavani and Ramachandran Ramjee and Alexey Tumanov},
	year         = 2024,
	url          = {
https://doi.org/10.48550/arXiv.2405.05465},
	eprint       = {2405.05465},
	archiveprefix = {arXiv},
	primaryclass = {cs.LG}
}

@misc{cho2024llmservingsimhwswcosimulationinfrastructure,
	title        = {LLMServingSim: A HW/SW Co-Simulation Infrastructure for LLM Inference Serving at Scale},
	author       = {Jaehong Cho and Minsu Kim and Hyunmin Choi and Guseul Heo and Jongse Park},
	year         = 2024,
	url          = {https://doi.org/10.48550/arXiv.2408.05499},
	eprint       = {2408.05499},
	archiveprefix = {arXiv},
	primaryclass = {cs.DC}
}

@misc{Chooseth12,
author = {},
  title = {Choose the k-NN algorithm for your billion-scale use case with OpenSearch | AWS Big Data Blog},
  url = "https://aws.amazon.com/blogs/big-data/choose-the-k-nn-algorithm-for-your-billion-scale-use-case-with-opensearch/",
month = {9},
year = {2022},
  note = "[Online; accessed 2025-04-11]"
}

@misc{dam2024chatbotsurvey,
      title={A Complete Survey on LLM-based AI Chatbots}, 
      author={Sumit Kumar Dam and Choong Seon Hong and Yu Qiao and Chaoning Zhang},
      year={2024},
      eprint={2406.16937},
      archivePrefix={arXiv},
      primaryClass={cs.CL},
      url={
https://doi.org/10.48550/arXiv.2406.16937
}, 
}

@misc{jiang2024codegen,
  title         = {A Survey on Large Language Models for Code Generation},
  author        = {Juyong Jiang and Fan Wang and Jiasi Shen and Sungju Kim and Sunghun Kim},
  year          = {2024},
  eprint        = {2406.00515},
  archivePrefix = {arXiv},
  primaryClass  = {cs.CL},
  url           = {https://doi.org/10.48550/arXiv.2406.00515},
  doi           = {10.48550/arXiv.2406.00515}
}

@misc{fiobenchmark_google,
  author = {{Google Cloud}},
  title = {{Benchmark Persistent Disk performance on a Linux VM}},
  howpublished = {\url{https://cloud.google.com/compute/docs/disks/benchmarking-pd-performance-linux}},
  note = {Last updated: 2025-08-07; Accessed: 2025-08-20},
  year = {2025}
}

@misc{gao2024ragsurvey,
  title         = {Retrieval-Augmented Generation for Large Language Models: A Survey},
  author        = {Yunfan Gao and Yun Xiong and Xinyu Gao and Kangxiang Jia and Jinliu Pan and Yuxi Bi and Yi Dai and Jiawei Sun and Meng Wang and Haofen Wang},
  year          = {2024},
  eprint        = {2312.10997},
  archivePrefix = {arXiv},
  primaryClass  = {cs.CL},
  url           = {https://doi.org/10.48550/arXiv.2312.10997},
  doi           = {10.48550/arXiv.2312.10997}
}

@misc{Etched,
	title        = {Etched is Making the Biggest Bet in AI},
	author       = {},
	year         = 2024,
	month        = {},
	url          = {https://www.etched.com/announcing-etched},
	note         = {(Accessed on 2/21/2025)}
}

@inproceedings{astrasim,
	title        = {{ASTRA-SIM}: Enabling SW/HW Co-Design Exploration for Distributed DL Training Platforms},
	author       = {Saeed Rashidi and Srinivas Sridharan and Sudarshan Srinivasan and Tushar Krishna},
	year         = 2020,
	booktitle    = {IEEE International Symposium on Performance Analysis of Systems and Software (ISPASS)},
    url = {https://www.doi.org/10.1109/ISPASS48437.2020.00018}
}

@misc{johnson2017billionscalesimilaritysearchgpus,
  title         = {Billion-scale similarity search with GPUs},
  author        = {Jeff Johnson and Matthijs Douze and Hervé Jégou},
  year          = {2017},
  eprint        = {1702.08734},
  archivePrefix = {arXiv},
  primaryClass  = {cs.CV},
  url           = {https://doi.org/10.48550/arXiv.1702.08734},
  doi           = {10.48550/arXiv.1702.08734}
}

@inproceedings{karpukhin2020densepassageretrievalopendomain,
  title     = {Dense Passage Retrieval for Open-Domain Question Answering},
  author    = {Vladimir Karpukhin and Barlas Oğuz and Sewon Min and Patrick Lewis and Ledell Wu and Sergey Edunov and Danqi Chen and Wen-tau Yih},
  booktitle = {Proceedings of the 2020 Conference on Empirical Methods in Natural Language Processing (EMNLP)},
  year      = {2020},
  address   = {Online},
  publisher = {Association for Computational Linguistics},
  pages     = {6769--6781},
  url       = {https://doi.org/10.18653/v1/2020.emnlp-main.550},
  doi       = {10.18653/v1/2020.emnlp-main.550}
}

@misc{hao_reasoning_2023,
  title        = {Reasoning with Language Model is Planning with World Model},
  author       = {Hao, Shibo and Gu, Yi and Ma, Haodi and Hong, Joshua Jiahua and Wang, Zhen and Wang, Daisy Zhe and Hu, Zhiting},
  year         = {2023},
  eprint       = {2305.14992},
  eprinttype   = {arXiv},
  doi          = {10.48550/arXiv.2305.14992},
  url          = {https://doi.org/10.48550/arXiv.2305.14992}
}

@misc{nvidia_ivf_pq,
	title        = {Accelerating Vector Search: NVIDIA cuVS IVF-PQ Part 1, Deep Dive | NVIDIA Technical Blog},
	author       = {Artem Chirkin},
	year         = 2024,
	month        = 7,
	url          = {https://developer.nvidia.com/blog/accelerating-vector-search-nvidia-cuvs-ivf-pq-deep-dive-part-1/},
	note         = {[Online; accessed 2025-03-08]}
}

@misc{bambhaniya2024demystifying,
  title         = {Demystifying AI Platform Design for Distributed Inference of Next-Generation LLM models},
  author        = {Abhimanyu Bambhaniya and Ritik Raj and Geonhwa Jeong and Souvik Kundu and Sudarshan Srinivasan and Suvinay Subramanian and Midhilesh Elavazhagan and Madhu Kumar and Tushar Krishna},
  year          = {2025},
  eprint        = {2406.01698},
  archivePrefix = {arXiv},
  primaryClass  = {cs.AR},
  url           = {https://doi.org/10.48550/arXiv.2406.01698},
  doi           = {10.48550/arXiv.2406.01698}
}

@misc{copilot,
	title        = {GitHub Copilot · Your AI pair programmer},
	author       = {MIcrosoft},
	url          = {https://github.com/features/copilot}
}

@misc{Ggerganov,
	title        = {GGERGANOV/llama.cpp: Port of facebook’s Llama model in C/C++},
	author       = {Ggerganov, Georgi},
	journal      = {GitHub},
	url          = {https://github.com/ggerganov/llama.cpp}
}

@article{holmes2024deepspeed,
	title        = {DeepSpeed-FastGen: High-throughput Text Generation for LLMs via MII and DeepSpeed-Inference},
	author       = {Holmes, Connor and Tanaka, Masahiro and Wyatt, Michael and Awan, Ammar Ahmad and Rasley, Jeff and Rajbhandari, Samyam and Aminabadi, Reza Yazdani and Qin, Heyang and Bakhtiari, Arash and Kurilenko, Lev and others},
	year         = 2024,
	url      = {
https://doi.org/10.48550/arXiv.2401.08671}
}

@inproceedings{kwon2023efficient,
	title        = {Efficient Memory Management for Large Language Model Serving with PagedAttention},
	author       = {Woosuk Kwon and Zhuohan Li and Siyuan Zhuang and Ying Sheng and Lianmin Zheng and Cody Hao Yu and Joseph E. Gonzalez and Hao Zhang and Ion Stoica},
	year         = 2023,
	booktitle    = {Proceedings of the ACM SIGOPS 29th Symposium on Operating Systems Principles},
    url  = {https://doi.org/10.48550/arXiv.2309.06180}
}

@misc{azurellmtrace2023,
  author       = {Microsoft Azure},
  title        = {{Azure Public Dataset: Azure LLM Inference Trace 2023}},
  year         = {2023},
  howpublished = {\url{https://github.com/Azure/AzurePublicDataset/blob/master/AzureLLMInferenceDataset2023.md}},
  note         = {Accessed: 2025-04-10}
}

@misc{splitwisesim2024,
  author       = {Mutinifni},
  title        = {{SplitwiseSim: LLM Serving Cluster Simulator}},
  year         = {2024},
  howpublished = {\url{https://github.com/Mutinifni/splitwise-sim}},
  note         = {Accessed: 2025-04-10}
}

@misc{vllm_splitwise2024,
  author       = {vLLM contributors},
  title        = {{Add Splitwise Implementation to vLLM}},
  year         = {2024},
  howpublished = {\url{https://github.com/vllm-project/vllm/pull/2809}},
  note         = {Accessed: 2025-04-10}
}

@inproceedings{pagedAttention,
	title        = {Efficient Memory Management for Large Language Model Serving with PagedAttention},
	author       = {Kwon, Woosuk and Li, Zhuohan and Zhuang, Siyuan and Sheng, Ying and Zheng, Lianmin and Yu, Cody Hao and Gonzalez, Joseph and Zhang, Hao and Stoica, Ion},
	year         = 2023,
	booktitle    = {Proceedings of the 29th Symposium on Operating Systems Principles},
	location     = {Koblenz, Germany},
	publisher    = {Association for Computing Machinery},
	address      = {New York, NY, USA},
	series       = {SOSP '23},
	pages        = {611–626},
	doi          = {10.1145/3600006.3613165},
	isbn         = 9798400702297,
	url          = {https://doi.org/10.1145/3600006.3613165},
	numpages     = 16
}

@misc{lightman2023letsverify,
  title         = {Let's Verify Step by Step},
  author        = {Hunter Lightman and Vineet Kosaraju and Yura Burda and Harri Edwards and Bowen Baker and Teddy Lee and Jan Leike and John Schulman and Ilya Sutskever and Karl Cobbe},
  year          = {2023},
  eprint        = {2305.20050},
  archivePrefix = {arXiv},
  primaryClass  = {cs.LG},
  url           = {https://doi.org/10.48550/arXiv.2305.20050},
  doi           = {10.48550/arXiv.2305.20050}
}

@misc{patel2023splitwise,
	title        = {Splitwise: Efficient generative LLM inference using phase splitting},
	author       = {Pratyush Patel and Esha Choukse and Chaojie Zhang and Íñigo Goiri and Aashaka Shah and Saeed Maleki and Ricardo Bianchini},
	year         = 2023,
	eprint       = {2311.18677},
	archiveprefix = {arXiv},
	primaryclass = {cs.AR},
    url = {https://doi.org/10.48550/arXiv.2311.18677}
}

@misc{TensorRT-LLM,
	title        = {NVIDIA. Tensorrt-llm.},
	author       = {NVIDIA},
	year         = 2023,
	url          = {https://github.com/NVIDIA/TensorRT-LLM}
}

@misc{triton,
	title        = {Triton TensorRT-LLM Backend},
	url          = {https://github.com/triton-inference-server/tensorrtllm\%5Fbackend}
}

@misc{qwen3technicalreport,
  title         = {Qwen3 Technical Report},
  author        = {Qwen Team},
  year          = {2025},
  eprint        = {2505.09388},
  archivePrefix = {arXiv},
  primaryClass  = {cs.CL},
  url           = {https://doi.org/10.48550/arXiv.2505.09388},
  doi           = {10.48550/arXiv.2505.09388}
}

@article{yao2024cacheblend,
  title   = {CacheBlend: Fast Large Language Model Serving with Cached Knowledge Fusion},
  author  = {Yao, Jiayi and Li, Hanchen and Liu, Yuhan and Ray, Siddhant and Cheng, Yihua and Zhang, Qizheng and Du, Kuntai and Lu, Shan and Jiang, Junchen},
  journal = {arXiv preprint arXiv:2405.16444},
  year    = {2024},
  url     = {https://doi.org/10.48550/arXiv.2405.16444},
  doi     = {10.48550/arXiv.2405.16444}
}

@inproceedings{yu2022orca,
  title     = {ORCA: A Distributed Serving System for {Transformer-Based} Generative Models},
  author    = {Yu, Gyeong-In and Jeong, Joo Seong and Kim, Geon-Woo and Kim, Soojeong and Chun, Byung-Gon},
  booktitle = {16th USENIX Symposium on Operating Systems Design and Implementation (OSDI 22)},
  year      = {2022},
  pages     = {521--538},
  publisher = {USENIX Association},
  url       = {https://www.usenix.org/conference/osdi22/presentation/yu}
}

@misc{hf_github-code-2025,
author = {Nick Saga},
  title = {nick007x/github-code-2025· Datasets at Hugging Face},
  url = "https://huggingface.co/datasets/nick007x/github-code-2025",
month = {10},
year = {2025},
  note = "[Online; accessed 2025-11-17]"
}

@article{kocisky-etal-2018-narrativeqa,
  title     = {The {N}arrative{QA} Reading Comprehension Challenge},
  author    = {Ko{\v{c}}isk{\'y}, Tom{\'a}{\v{s}} and
               Schwarz, Jonathan and
               Blunsom, Phil and
               Dyer, Chris and
               Hermann, Karl Moritz and
               Melis, G{\'a}bor and
               Grefenstette, Edward},
  journal   = {Transactions of the Association for Computational Linguistics},
  volume    = {6},
  year      = {2018},
  address   = {Cambridge, MA},
  publisher = {MIT Press},
  pages     = {317--328},
  doi       = {10.1162/tacl_a_00023},
  url       = {https://doi.org/10.1162/tacl_a_00023}
}

@article{zhang2023hardware,
  title   = {A Hardware Evaluation Framework for Large Language Model Inference},
  author  = {Zhang, Hengrui and Ning, August and Prabhakar, Rohan and Wentzlaff, David},
  year    = {2023},
  journal = {arXiv preprint arXiv:2312.03134},
  url     = {https://doi.org/10.48550/arXiv.2312.03134},
  doi     = {10.48550/arXiv.2312.03134}
}

@misc{zhong2024distserve,
  title         = {DistServe: Disaggregating Prefill and Decoding for Goodput-optimized Large Language Model Serving},
  author        = {Yinmin Zhong and Shengyu Liu and Junda Chen and Jianbo Hu and Yibo Zhu and Xuanzhe Liu and Xin Jin and Hao Zhang},
  year          = {2024},
  eprint        = {2401.09670},
  archivePrefix = {arXiv},
  primaryClass  = {cs.DC},
  url           = {https://doi.org/10.48550/arXiv.2401.09670},
  doi           = {10.48550/arXiv.2401.09670}
}

@misc{agrawal2024taming,
      title={Taming Throughput-Latency Tradeoff in LLM Inference with Sarathi-Serve}, 
      author={Amey Agrawal and Nitin Kedia and Ashish Panwar and Jayashree Mohan and Nipun Kwatra and Bhargav S. Gulavani and Alexey Tumanov and Ramachandran Ramjee},
      year={2024},
      eprint={2403.02310},
      archivePrefix={arXiv},
      primaryClass={cs.LG},
      url={
https://doi.org/10.48550/arXiv.2403.02310}, 
}

@misc{GitHubNV14:online,
	title        = {GitHub - NVIDIA/FasterTransformer: Transformer related optimization, including BERT, GPT},
	author       = {NVIDIA},
	year         = {},
	month        = {},
	url          = {https://github.com/NVIDIA/FasterTransformer},
	note         = {[Online; accessed 2025-04-10]}
}

@misc{nv_dynamo,
author = {},
  title = {Dynamo Inference Framework | NVIDIA Developer},
  url = "https://developer.nvidia.com/dynamo",
month = {},
year = {},
  note = "[Online; accessed 2025-04-10]"
}

@inproceedings{liu2024cachegen,
  title={Cachegen: Kv cache compression and streaming for fast large language model serving},
  author={Liu, Yuhan and Li, Hanchen and Cheng, Yihua and Ray, Siddhant and Huang, Yuyang and Zhang, Qizheng and Du, Kuntai and Yao, Jiayi and Lu, Shan and Ananthanarayanan, Ganesh and others},
  booktitle={Proceedings of the ACM SIGCOMM 2024 Conference},
  pages={38--56},
  year={2024},
  url = {https://doi.org/10.48550/arXiv.2310.07240}
}

@article{cheng2024large,
  title={Do Large Language Models Need a Content Delivery Network?},
  author={Cheng, Yihua and Du, Kuntai and Yao, Jiayi and Jiang, Junchen},
  url={
https://doi.org/10.48550/arXiv.2409.13761},
  year={2024}
}

@misc{jain2025intelligentrouterllmworkloads,
  title         = {Intelligent Router for LLM Workloads: Improving Performance Through Workload-Aware Load Balancing},
  author        = {Kunal Jain and Anjaly Parayil and Ankur Mallick and Esha Choukse and Xiaoting Qin and Jue Zhang and Íñigo Goiri and Rujia Wang and Chetan Bansal and Victor Rühle and Anoop Kulkarni and Steve Kofsky and Saravan Rajmohan},
  year          = {2025},
  eprint        = {2408.13510},
  archivePrefix = {arXiv},
  primaryClass  = {cs.DC},
  url           = {https://doi.org/10.48550/arXiv.2408.13510},
  doi           = {10.48550/arXiv.2408.13510}
}

@misc{astrasim_doc_ns3,
author = {Astra-Sim},
  title = {ns-3 Network Backend — ASTRA-sim 2.2 documentation},
  url = "https://astra-sim.github.io/astra-sim-docs/network-backend/ns3-network-backend.html",
month = {},
year = {},
  note = "[Online; accessed 2025-06-19]"
}

@misc{moduler_prefix_caching,
author = {},
  title = {What is Prefix Caching? A Beginner's Guide - AI Resources},
  url = "https://www.modular.com/ai-resources/what-is-prefix-caching-a-beginner-s-guide",
month = {},
year = {},
  note = "[Online; accessed 2025-04-11]"
}

@misc{vllm_apc,
author = {},
  title = {Automatic Prefix Caching — vLLM},
  url = "https://docs.vllm.ai/en/latest/design/v1/prefix_caching.html",
month = {},
year = {},
  note = "[Online; accessed 2025-04-11]"
}
